\documentclass[11pt]{article}

\usepackage{fullpage}
\usepackage{comment}
\usepackage{ifthen}
\usepackage{amsthm}
\usepackage{tikz}
\usetikzlibrary{positioning,calc,arrows.meta}
\usepackage{pgfplots}
\pgfplotsset{compat=1.18}
\usetikzlibrary{patterns}
\usepgfplotslibrary{fillbetween}
\usetikzlibrary{intersections}
\usepackage{csquotes}

\usepackage[dvipsnames]{xcolor}
\usepackage{bigints}

\usepackage{amsmath,amssymb}
\usepackage{xcolor}
\usepackage{tcolorbox}

\usepackage{thm-restate}

\usepackage{natbib}
\bibpunct{(}{)}{;}{a}{,}{,}

\usepackage{hyperref}       

\usepackage{comment}
\newcommand{\xhdr}[1]{\par\smallskip\noindent\textbf{#1}\ }

\renewcommand{\paragraph}{\xhdr}

\usepackage{url}            
\usepackage{booktabs}       
\usepackage{amsfonts}       
\usepackage{nicefrac}       
\usepackage{microtype}      
\usepackage{dsfont}

\usepackage{mathtools, bm}

\usepackage{algorithm}
\usepackage{algpseudocode}

\usepackage{subcaption}
\usepackage{multirow}
\usepackage{float}
\usepackage{xspace}

\usepackage{enumitem}

\usepackage{xfrac}

\usepackage{cleveref}
\crefname{enumi}{part}{parts}
\crefname{equation}{eq.}{eqs.}

\theoremstyle{plain}
\newtheorem{theorem}{Theorem}[section]
\newtheorem{lemma}[theorem]{Lemma}

\newtheorem{proposition}[theorem]{Proposition}

\theoremstyle{plain}
\newtheorem{definition}{Definition}[section] 
\newtheorem{example}[definition]{Example}

\usepackage{color-edits}
\addauthor[Yiding]{yf}{purple}    
\addauthor[Mengfan]{mf}{blue}
\addauthor[Yaohao]{yh}{brown}
\addauthor[Tian]{tb}{Maroon}

\allowdisplaybreaks
\usepackage{etoolbox}
\preto\part{\setcounter{section}{0}}

\newcommand{\val}{\theta}                
\newcommand{\valdis}{\mathcal{F}}        
\newcommand{\valdens}{\valdis'}          
\newcommand{\inversevaldis}{\valdis^{-1}}
\newcommand{\valL}{\underline{\theta}}   
\newcommand{\valU}{\overline{\theta}}    

\newcommand{\costpar}{\mu}               
\newcommand{\costdis}{\mathcal{G}}       
\newcommand{\inversecostdis}{\costdis^{-1}}
\newcommand{\costdens}{\costdis'}        
\newcommand{\costL}{\underline{\mu}}     
\newcommand{\costU}{\overline{\mu}}      

\newcommand{\qualities}{Q}               
\newcommand{\effort}{c}                  
\newcommand{\slope}{\kappa}                

\newcommand{\effortgap}{\Delta\effort}  

\newcommand{\posted}{T}                  
\newcommand{\menu}{M}                    
\newcommand{\tiers}{n}                   
\newcommand{\convextiers}{L}
\newcommand{\menusize}{m}                

\newcommand{\quality}{q}                 
\newcommand{\qTier}[1]{q^{(#1)}}         

\newcommand{\qualitygap}{\Delta\quality}  

\newcommand{\price}{p}                   

\newcommand{\wage}{w}                    

\newcommand{\qC}{q_{\mathrm{C}}}         
\newcommand{\qP}{q_{\mathrm{P}}}         
\newcommand{\Ucons}{U_{\mathrm{C}}}      
\newcommand{\Uprov}{U_{\mathrm{P}}}      

\newcommand{\demdist}{\mathcal{D}}       
\newcommand{\supdist}{\mathcal{S}}       
\newcommand{\tradedist}{\mathcal{T}}     

\newcommand{\profit}{\Pi}                
\newcommand{\profitctrib}{\pi}

\newcommand{\welfare}{\mathsf{W}}        
\newcommand{\welfarectrib}{\omega}

\newcommand{\var}{z}                    

\newcommand{\profitopt}{\var^{\profit}}            

\newcommand{\welfareopt}{\var^{W}}

\newcommand{\parameter}{\xi}
\newcommand{\approxrate}{\varepsilon}
\newcommand{\discresize}{\delta}

\newcommand{\commonfunc}{J}

\newcommand{\valhrate}{h}
\newcommand{\costhrate}{r}
\newcommand{\hazard}{H}

\newcommand{\oprofit}{\mathrm{OPT}}
\newcommand{\oracle}{Z}
\newcommand{\oraclevalue}{R}

\newcommand{\dpfunc}{D}                  
\newcommand{\normconst}{S}               

\newcommand{\welfareratio}{\Gamma}
\newcommand{\qualityU}{\overline{\quality}}
\newcommand{\convexeffort}{\underline{\effort}}
\newcommand{\costgap}{d}                 
\newcommand{\convexslope}{\convexeffort'}         
\newcommand{\qualityvar}{s}
\newcommand{\massvar}{u}
\newcommand{\logmass}{\xi}               
\newcommand{\tierindex}{t}
\newcommand{\sumindex}{\tau}

\newcommand{\leftindex}{i}
\newcommand{\rightindex}{j}
\newcommand{\groupcount}{k}

\newcommand{\testmass}{y}               
\newcommand{\masssmall}{x}
\newcommand{\masslarge}{y}
\newcommand{\massP}{\testmass^{P}}       
\newcommand{\massW}{\testmass^{W}}       

\newcommand{\consumermean}{\mathbb{E}[\val]}

\newcommand{\typevar}{\nu}
\newcommand{\truncation}{K}
\newcommand{\approxstage}{N}
\newcommand{\ee}{{\mathrm e}}

\newcommand{\parent}[1]{\left(#1\right)}

\newcommand{\deq}{\triangleq}

\DeclareMathOperator*{\argmax}{arg\,max}

\newcommand{\reals}{\mathbb{R}}

\newcommand{\naturals}{\mathbb{N}}

\DeclareMathOperator*{\prb}{\mathrm{Pr}}
\newcommand{\prob}[2][]{\prb\ifthenelse{\not\equal{}{#1}}{\nolimits_{#1}}{}\!\left[{\def\givenn{\middle|}#2}\right]}
\newcommand{\expect}[2][]{\mathbb{E}\ifthenelse{\not\equal{}{#1}}{_{#1}}{}\!\left[{\def\givenn{\middle|}#2}\right]}

\newcommand{\tparen}{\big}
\newcommand{\tprob}[2][]{\text{Pr}\ifthenelse{\not\equal{}{#1}}{_{#1}}{}\tparen[{\def\given{\tparen|}#2}\tparen]}
\newcommand{\texpect}[2][]{\mathbb{E}\ifthenelse{\not\equal{}{#1}}{_{#1}}{}\tparen[{\def\given{\tparen|}#2}\tparen]}

\newcommand{\sprob}[2][]{\text{Pr}\ifthenelse{\not\equal{}{#1}}{_{#1}}{}[#2]}
\newcommand{\sexpect}[2][]{\mathbb{E}\ifthenelse{\not\equal{}{#1}}{_{#1}}{}[#2]}

\newcommand{\dd}{{\mathrm d}}

\newcommand{\bO}[1]{\mathcal{O}\parent{#1}}

\newcommand{\indicator}[1]{{\mathbbm{1}\left\{ #1 \right\}}}

\providecommand{\mathbbm}[1]{\mathds{#1}}

\author{
Yaohao Liu\thanks{University of Electronic Science and Technology of China. Email: {\tt yaohao.liu594@gmail.com}}
\and
Mengfan Ma\thanks{Central China Normal University. Email: {\tt mengfanma1@gmail.com}}
\and
Mingyu Xiao\thanks{University of Electronic Science and Technology of China. Email: {\tt myxiao@uestc.edu.cn}}
}

\begin{document}
\title{Two-Sided Screening in Quality-Differentiated Markets}

\maketitle

\begin{abstract}
    

Motivated by digital marketplaces such as Uber and Airbnb, we study how a two-sided platform designs a menu of qualities, consumer prices, and provider wages.
Consumers and providers self-select according to their private values and costs, so the platform must design its menu to attract consumers while inducing enough supply to meet their demand.
We study this coordination problem over a continuous quality space, requiring supply to cover demand at every quality threshold.

We formulate menu design as an optimization over trading distributions, which specify how much trade occurs at or above each quality.
Through convexification of the technological cost function, we characterize the profit- and welfare-optimal trading distributions.
This characterization shows how production technology alone can reduce menu size without sacrificing optimal profit or welfare, regardless of the type distributions, and determines when optimal menus can be finite.
We then show that profit maximization can be highly efficient: under monotone hazard rate conditions on both sides, a profit-optimal menu achieves at least $1/(\ee-1)$ of optimal social welfare.

We next consider the practical setting in which the menu size is restricted to a prescribed number of offers.
We establish how the restricted menu approximates the unrestricted optimum in profit.
We develop an algorithm that computes a menu within this size limit in polynomial time.
The computed menu simultaneously approximates the unrestricted optimal profit and social welfare with additive error.

\end{abstract}

\section{Introduction}
\label{sec:introduction}
Platforms in quality-differentiated markets coordinate trade in goods or services offered at different quality levels.
For example, Uber offers service tiers such as UberX and Uber Comfort.
Offering a higher-quality option requires both consumers willing to pay for it and providers willing to supply it at the offered compensation.
This creates a joint design problem: consumer prices shape demand across qualities, while provider compensations shape the corresponding supply.
We study how a platform can coordinate two-sided menus when supply must cover demand at every quality, and how its choices affect platform profit and social welfare.

This problem combines screening with a feasibility constraint that links the two sides.
In classical one-sided screening, a seller offers a price--quality menu to consumers with privately known valuations \citep{MR-78}.
There, the seller produces whatever quality a consumer selects, so every consumer's choice is fulfilled automatically.
Here, the platform also faces privately informed providers and must design a compensation menu to induce their participation and quality choices.
Adjusting either menu changes the distribution of demand or supply across qualities, so the two menus cannot be optimized separately.
Moreover, a continuous quality space may require an infinite optimal menu, while practical considerations favor finite menus that are easier to communicate and manage.
These considerations raise three related questions:
\par\smallskip
\noindent(1) \emph{What structure do optimal two-sided menus have?}
\par
\noindent(2) \emph{How much welfare does profit maximization preserve?}
\par
\noindent(3) \emph{How much profit can a platform retain with a limited menu, and how can such a menu be computed?}
\par\smallskip

To address these questions, we study a platform intermediating trade between a continuum of consumers and a continuum of providers over a compact quality interval.
Consumers have private values for quality, and providers have private cost types that scale a common technological cost function.
The platform posts a set of qualities with consumer prices and provider compensations.
Traders on both sides independently select their preferred quality or choose not to participate.
We require supply to cover demand at every quality, with excess supply downgraded or discarded at no cost, and show that optimal menus match the two exactly (\Cref{prop:strict_matching}).
Our goal is to characterize optimal menus in quality-differentiated markets, quantify the welfare consequences of profit maximization, and develop profit and welfare guarantees for menus with a limited number of qualities.

\subsection{Our contribution and techniques}
Within our model, we address the structure of optimal menus, the welfare consequences of profit maximization, and the design of limited menus.
Below we present our main contributions with a running example (\Cref{ex:running}) for illustrating.

\begin{example}\label{ex:running}
Suppose the platform chooses qualities from the interval $[0,1]$. 
Both consumer and provider types are uniformly distributed on $[0,1]$, and the technological cost function is
\begin{align*}
    \effort(\quality)=
    \begin{cases}
        \frac{3}{2}\quality-\quality^{2}, & 0\leq\quality\leq\frac12,\\
        \frac{9}{2}\quality-\quality^{2}-\frac{3}{2}, & \frac12<\quality\leq1.
    \end{cases}
\end{align*}
\end{example}

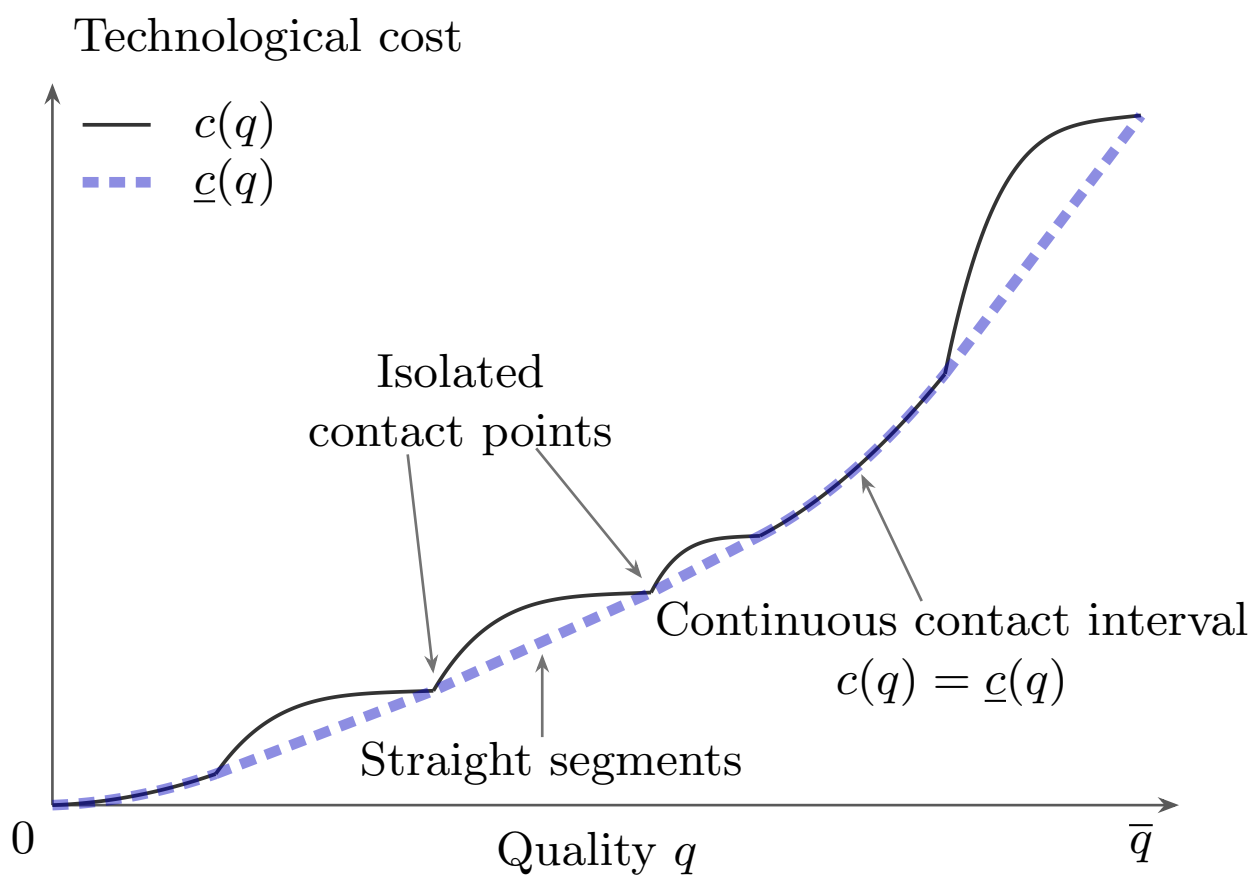
\begin{figure}[t]
    \centering
    \begin{tikzpicture}[
    font=\footnotesize,
    axis/.style={draw=black!65, line width=0.5pt, -{Stealth[length=4pt]}},
    original/.style={draw=black!80, line width=0.7pt},
    minorant/.style={draw=blue!75!black, line width=2pt, draw opacity=0.45, dash pattern=on 3pt off 2pt},
    pointer/.style={draw=black!55, line width=0.5pt, -{Stealth[length=3.5pt]}},
    annotation/.style={align=center, inner sep=2pt}
]
    \node[anchor=west] at (0,5.07) {Technological cost};
    \draw[original] (0.2,4.5) -- (0.65,4.5);
    \node[anchor=west] at (0.8,4.5) {$\effort(\quality)$};
    \draw[minorant] (0.2,4.12) -- (0.65,4.12);
    \node[anchor=west] at (0.8,4.12) {$\convexeffort(\quality)$};

    \begin{scope}[x=0.72cm,y=0.25cm]
        \draw[axis] (0,0) -- (10.35,0);
        \draw[axis] (0,0) -- (0,19.1);

        \draw[original,domain=0:1.5,samples=45]
            plot (\x,{0.1*\x+0.3*\x*\x});
        \draw[original,domain=1.5:3.5,samples=65]
            plot (\x,{0.825+2.2*(0.08*((\x-1.5)/2)+0.92*(1-(1-(\x-1.5)/2)^4))});
        \draw[original,domain=3.5:5.5,samples=65]
            plot (\x,{3.025+2.6*(0.08*((\x-3.5)/2)+0.92*(1-(1-(\x-3.5)/2)^4))});
        \draw[original,domain=5.5:6.5,samples=45]
            plot (\x,{5.625+1.5*(0.08*(\x-5.5)+0.92*(1-(1-(\x-5.5))^4))});
        \draw[original,domain=6.5:8.2,samples=65]
            plot (\x,{7.125+1.5*(\x-6.5)+0.6*(\x-6.5)^2});
        \draw[original,domain=8.2:10,samples=65]
            plot (\x,{11.409+6.84*(0.08*((\x-8.2)/1.8)+0.92*(1-(1-(\x-8.2)/1.8)^4))});

        \draw[minorant,domain=0:1.5,samples=45]
            plot (\x,{0.1*\x+0.3*\x*\x});
        \draw[minorant] (1.5,0.825) -- (3.5,3.025) -- (5.5,5.625)
            -- (6.5,7.125);
        \draw[minorant,domain=6.5:8.2,samples=65]
            plot (\x,{7.125+1.5*(\x-6.5)+0.6*(\x-6.5)^2});
        \draw[minorant] (8.2,11.409) -- (10,18.249);

        \node[below left,inner sep=3pt] at (0,0) {$0$};
        \node[below,inner sep=3pt] at (10,0) {$\qualityU$};

    \end{scope}

    \node[annotation] at (2.7,2.65) {Isolated\\contact points};
    \draw[pointer] (2.35,2.32) -- (2.52,0.87);
    \draw[pointer] (3.2,2.36) -- (3.9,1.5);
    \node[annotation] at (3.3,0.28) {Straight segments};
    \draw[pointer] (3.24,0.45) -- (3.24,1.0);
    \node[annotation] at (5.95,1.0)
        {Continuous contact interval\\$\effort(\quality)=\convexeffort(\quality)$};
    \draw[pointer] (5.75,1.35) -- (5.35,2.19);
    \node at (3.6,-0.32) {Quality $\quality$};
\end{tikzpicture}
    \caption{Convexification of technological cost.}
    \label{fig:common_convexification}
\end{figure}

\xhdr{Structure of optimal menus.}
We formulate menu design as an optimization over a common trading distribution, which specify how much trade occurs at or above each quality and express platform profit and social welfare as integrals of scalar contributions (\Cref{thm:characterization}).
Our key reduction replaces technological cost function $\effort$ by its greatest convex minorant $\convexeffort$, the largest convex function below it (\Cref{thm:commonconvexification}).
As illustrated in \Cref{fig:common_convexification}, the solid and dashed curves represent $\effort$ and $\convexeffort$. 
They coincide at isolated contact qualities and along contact intervals, while $\convexeffort$ is linear between contacts wherever the curves differ.
This replacement preserves both optimal values, and each objective admits an optimal menu using only contact qualities.
Thus, technology alone identifies qualities that can be omitted, while type distributions determine the optimal trading masses.
We also give a necessary and sufficient condition for finite optimal menus (\Cref{prop:finite_optimal_menu}).
\Cref{ex:running_structure} below illustrates the trading-distribution representation, cost convexification, and the resulting finite optimal menu.

\addtocounter{example}{-1}
\begin{example}[cont.]
\phantomsection
\label{ex:running_structure}
Write $\var(\quality)$ for the mass trading at quality at least $\quality$.
The strictly matched menu we construct to implement $\var(\quality)$ has profit $\int_0^1\bigl[\var(\quality)-(1+\effort'(\quality))\var(\quality)^2\bigr]\,d\quality$ and welfare $\int_0^1\bigl[\var(\quality)-\tfrac12(1+\effort'(\quality))\var(\quality)^2\bigr]\,d\quality$.
The greatest convex minorant joins the contact points $(0,0)$, $(1/2,1/2)$, and $(1,2)$:
\begin{align*}
    \convexeffort(\quality)=
    \begin{cases}
        \quality, & 0\leq\quality\leq\frac12,\\
        3\quality-1, & \frac12<\quality\leq1.
    \end{cases}
\end{align*}

The profit-maximizing menu is finite: it offers quality $1/2$ at price $3/8$ and compensation $1/8$, and quality $1$ at price $13/16$ and compensation $5/16$, plus the outside option with optimal profit $3/32$.

The welfare optimum is also attained by a finite menu offering the same two qualities with optimal welfare $3/16$.
\end{example}

\xhdr{Efficiency of profit maximization.}
We characterize the efficiency loss caused by profit maximization.
The welfare optimum can be implemented without a platform subsidy (\Cref{prop:weak_budget_balance}), yet profit maximization can preserve arbitrarily little welfare even when both type distributions are regular (\Cref{prop:regular_insufficient}).
Under MHR condition, we establish a tight welfare ratio $1/(\ee-1)$ (\Cref{thm:mhr_ratio}).
\Cref{ex:running_efficiency} below illustrates the welfare comparison between profit and welfare maximization under MHR.

\addtocounter{example}{-1}
\begin{example}[cont.]
\phantomsection
\label{ex:running_efficiency}
Both uniform type distributions satisfy MHR conditions.
The profit-maximizing menu yields welfare $9/64$ and therefore preserves $3/4$ of optimal welfare $3/16$.
This ratio exceeds $1/(\ee-1)$, consistent with our welfare guarantee.
\end{example}

\xhdr{Limited menus and computation.}
When the smallest unrestricted optimal menu uses $\tiers$ qualities, restricting menus to at most $\menusize\leq\tiers$ qualities preserves at least a $\menusize/\tiers$ fraction of optimal profit, and this bound is tight (\Cref{thm:restrictmenu_profitloss}).
For general instances, including those requiring infinite optimal menus, we develop a dynamic program under a scalar profit oracle (\Cref{thm:compute_restricted_menu}).
The algorithm combines quality discretization, cost convexification, and optimization over consecutive groups of quality increments.
It returns a menu with at most $\menusize$ qualities and additive profit loss at most $\qualityU \cdot \consumermean/\menusize+\approxrate$, where $\qualityU$ is the quality bound, $\consumermean$ is the mean consumer value, and $\approxrate>0$ is an accuracy parameter.
Under MHR, the same output achieves at least $1/\ee$ of optimal welfare minus the same additive error.
\Cref{ex:running_computation} below illustrates the computed restricted menu and compares its profit and welfare with our approximation guarantees.

\addtocounter{example}{-1}
\begin{example}[cont.]
\phantomsection
\label{ex:running_computation}
Set $\menusize=1$ and $\approxrate=1/4$.
The algorithm returns a menu offering only quality $1$ at price $5/6$ and compensation $1/3$, in addition to the outside option.
Its profit is $1/12$, retaining $8/9$ of the unrestricted optimal profit and losing only $1/96$.
Its welfare is $1/8$, retaining $2/3$ of optimal welfare.

The retained profit exceeds our $1/2$ relative guarantee, while the profit and welfare also satisfy the general additive bounds.
\end{example}

\subsection{Related work}

Our model builds on monopolistic price-quality screening \citep{MR-78} and its multidimensional extensions \citep{RC-98,AR-99}.
We study private information on both sides of a platform, with provider choices required to cover consumer choices at every quality.
Similar supply--demand matching conditions are common in models of quality-differentiated markets and platforms \citep[e.g.,][]{E-10,LJW-24}.

\xhdr{Matching intermediation.}
A broad literature studies pricing, screening, and matching by intermediaries \citep{DL-07,J-13,GP-16,YA-22,FP-22,GP-24,BFLMX-25}.
Algorithmic work studies intermediary fees \citep{NYK-14}, revenue-maximizing double auctions \citep{DGTZ-14}, and match recommendations \citep{AKMSS-20}.
Related work examines matching and assignment under distributional constraints \citep{KK-15,ASS-19}.
We focus on consumer and provider menus over common production qualities, requiring the induced supply to cover the induced demand at every quality.

\xhdr{Menu structure and pooling.}
Prior work studies optimal nested bundles \citep{BBHS-21}, pooling under cross-side externalities \citep{JKM-22}, and information design concerning buyer values \citep{BHM-26}.
Algorithmic studies also examine optimal information disclosure \citep{DX-16} and its limits under communication constraints \citep{DKQ-16}.
\citet{LJW-24} also require supply to meet demand on a platform whose buyers value quality linearly, but seller quality is exogenous and the platform chooses which sellers to admit and what quality information to disclose.
They show that, under a demand-elasticity condition, admitting only high-quality sellers and pooling them maximizes revenue.
In our model, providers produce quality at privately known costs and the platform sets transfers on both sides.
Our structural reduction instead convexifies technological cost independently of type distributions, preserving both optimal profit and optimal welfare.

\xhdr{Profit and gains from trade.}
Existing work examines matching distortions under profit maximization \citep{GP-16,YA-22} and designs mechanisms approximating second-best \citep{BCWZ-17} or first-best \citep{DMSW-22} gains from trade.
Other work bounds revenue relative to welfare in single-parameter environments \citep{KY-13} and approximates gains from trade in multidimensional markets \citep{CGMZ-21}.
We establish a tight welfare guarantee for profit-maximizing quality menus under MHR.

\xhdr{Limited menus and computation.}
\citet{BYZ-21} quantify finite-menu losses in nonlinear pricing.
\citet{BGN-17} bound the menu size needed to approximate revenue for an additive buyer with independent item values.
Further work studies menu complexity with private values and deadlines \citep{SSW-18} and computational hardness in multi-item mechanism design \citep{DDT-14}.
We study a common quality limit on both sides, obtaining a dynamic program and approximation guarantees.

\section{Preliminaries}
\label{sec:preliminary}
We consider a market that consists of two sides, consumers and providers, and a monopolistic trading platform.
On one side is a unit mass of providers, each of whom can produce one unit of a service or commodity at different quality levels.
Denote by $\qualities\deq[0,\qualityU]$ the quality set that can be produced by providers.
The other side is a unit mass of consumers, each with unit demand for the service or commodity from the providers.
Consumers and providers cannot trade with each other directly and every transaction takes place through the platform.


\xhdr{Provider and consumer types.}
On the provider side, the production of quality $\quality$ requires an intrinsic effort $\effort(\quality)$, where the technological cost function $\effort:\qualities\to[0,\infty)$ is absolutely continuous and strictly increasing with $\effort(0)=0$.
A provider's type $\costpar$ is her cost per unit of effort, so her cost of producing quality $\quality$ is $\costpar\cdot\effort(\quality)$.
Suppose the compensation for producing quality $\quality$ is $\wage(\quality)$.
Then the utility of a provider of type $\costpar$ who produces quality $\quality$ is $\wage(\quality)-\costpar\cdot\effort(\quality)$.
The unit mass of providers is distributed over types according to the cumulative distribution $\costdis$ with density $\costdens$, supported on the interval $[\costL,\costU]$ with $0\leq\costL<\costU<\infty$.
That is, the mass of providers with types at most $\costpar$ is $\costdis(\costpar)$ for every $\costpar\in[\costL,\costU]$.

The consumer's type $\val$ is the consumer's marginal value of quality.
Suppose the price for buying quality $\quality$ is $\price(\quality)$.
Then the utility of a consumer of type $\val$ who buys quality $\quality$ is $\val\cdot\quality-\price(\quality)$.
The consumer's value type $\val$ has cumulative distribution $\valdis$ and density $\valdens$, supported on the interval $[\valL,\valU]$ with $0\leq\valL<\valU<\infty$.
Both type distributions are strictly increasing on their supports.
For the type distributions $\valdis$ and $\costdis$, let the corresponding \emph{quantile functions} be $\inversevaldis:[0,1]\to[\valL,\valU]$ and $\inversecostdis:[0,1]\to[\costL,\costU]$, mapping each mass of traders to the type below which that mass lies.
\xhdr{Two-sided screening.}
The platform knows the \emph{instance} $(\qualities,\effort,\valdis,\costdis)$, consisting of the quality set, the technological cost function, and the two type distributions, but types are traders' private information.
Hence, the platform cannot tailor prices or wages to individual traders; instead, it commits to a two-sided menu of take-it-or-leave-it offers, from which each trader selects one.
The menu thereby \emph{screens} traders by self-selection, sorting each side along the quality spectrum according to type.
\begin{definition}[Two-sided menu]\label{def:menu}
A \emph{two-sided menu} is a tuple $\menu=(\posted,\price,\wage)$ consisting of a posted quality set $\posted\subseteq\qualities$, a consumer price schedule $\price:\posted\to\reals_{\geq0}$, and a provider wage schedule $\wage:\posted\to\reals_{\geq0}$.
Moreover, the menu posts the outside option at zero transfers: $0\in\posted$ and $\price(0)=\wage(0)=0$.
An \emph{offer} $(\quality,\price(\quality),\wage(\quality))$ of the menu $\menu$ is a posted quality $\quality\in\posted$ together with its price and wage.
\end{definition}
Note that the posted quality set $\posted$ need not be the whole quality set $\qualities$.
Given a menu $\menu=(\posted,\price,\wage)$, a consumer of type $\val$ selects a posted quality $\qC(\val)$ that maximizes her utility,\footnote{We consider menus under which every type has a utility-maximizing posted quality.} and a provider of type $\costpar$ selects a posted quality $\qP(\costpar)$ that maximizes hers:
\begin{align*}
\qC(\val)\in\argmax_{\quality\in\posted}\{\val\cdot\quality-\price(\quality)\},\quad
\qP(\costpar)\in\argmax_{\quality\in\posted}\{\wage(\quality)-\costpar\cdot\effort(\quality)\}.
\end{align*}
We call $\qC:[\valL,\valU]\to\posted$ and $\qP:[\costL,\costU]\to\posted$ the consumer and provider \emph{choice maps}, with ties broken by a fixed rule when there is more than one maximizer.
Thus, the type-$\val$ consumer's and type-$\costpar$ provider's \emph{indirect utilities} are defined by
\begin{align*}
\Ucons(\val)\deq\sup\nolimits_{\quality\in\posted}\{\val\cdot\quality-\price(\quality)\},\quad\Uprov(\costpar)\deq\sup\nolimits_{\quality\in\posted}\{\wage(\quality)-\costpar\cdot\effort(\quality)\}.
\end{align*}

\xhdr{Demand, supply and market clearing.}
Given an instance $(\qualities,\effort,\valdis,\costdis)$ and a menu $\menu$, the choice maps determine the distribution of selected qualities on each side of the market, which we call the \emph{demand and supply distributions}, defined as follows:
\begin{definition}[Demand and supply distributions]\label{def:demand_supply}
For an instance $(\qualities,\effort,\valdis,\costdis)$ and a menu $\menu=(\posted,\price,\wage)$, the induced \emph{demand distribution} $\demdist$ and \emph{supply distribution} $\supdist$ give, for each quality $\quality$, the masses of consumers and providers who select a quality of at least $\quality$:
\begin{align*}
\demdist(\quality)\deq\prob[\val]{\qC(\val)\geq\quality},\quad\supdist(\quality)\deq\prob[\costpar]{\qP(\costpar)\geq\quality}.
\end{align*}
\end{definition}

Ideally, we should have $\demdist=\supdist$ so that every provider's output is consumed by a consumer who selects exactly that quality.
We call this condition \emph{strict matching}.
More generally, service can be downgraded or discarded at no cost, so a consumer of type $\val$ can be served with any quality equal to or higher than $\qC(\val)$ while still paying $\price(\qC(\val))$.
A provider of type $\costpar$ still produces quality $\qP(\costpar)$ for the wage $\wage(\qP(\costpar))$, even though her output may be downgraded or discarded.
All consumers can be served exactly when $\demdist(\quality)\leq\supdist(\quality)$ at every quality $\quality$, that is, when the providers who produce quality $\quality$ or higher are enough to serve the consumers who select quality $\quality$ or higher, a condition we call \emph{feasibility}.
Market clearing can then be achieved by pairing both sides from the highest quality down.
Formally, we have the following definition.
\begin{definition}[Feasible and strict matching]\label{def:strict_matching}
For an instance $(\qualities,\effort,\valdis,\costdis)$, a menu $\menu$ is \emph{feasible} if the induced demand distribution $\demdist$ and supply distribution $\supdist$ satisfy $\demdist(\quality)\leq\supdist(\quality)$ for every positive quality $\quality$.
A menu $\menu$ satisfies \emph{strict matching} if $\demdist=\supdist$, and we call the common distribution the \emph{trading distribution}, denoted by $\tradedist$.
\end{definition}

For a strictly matched menu $\menu$, the trading distribution $\tradedist:[0,\qualityU]\to[0,1]$ is nonincreasing and left-continuous.
We call $\tradedist(\quality)$ the \emph{trading mass} at $\quality$, which is the common mass of consumers and providers who trade at a quality at least $\quality$.
\xhdr{Objectives.}
The platform's primary objective is profit. The profit of a feasible menu $\menu$ is the expected consumer payment minus the expected provider compensation,
\begin{align}
\profit\deq\expect[\val]{\price(\qC(\val))}-\expect[\costpar]{\wage(\qP(\costpar))}.\label{eq:profit}
\end{align}
We also consider social welfare as an efficiency benchmark.
The social welfare of a feasible menu $\menu$ is the expected consumer gross value minus the expected provider production cost,
\begin{align}
\welfare\deq\expect[\val]{\val\cdot\qC(\val)}-\expect[\costpar]{\costpar\cdot\effort(\qP(\costpar))}.\label{eq:welfare}
\end{align}
Both objectives are maximized over feasible menus.
We conclude this section by showing that focusing on strictly matched menus is without loss of generality.
\begin{proposition}\label{prop:strict_matching}
For an instance $(\qualities,\effort,\valdis,\costdis)$, every feasible menu that is not strictly matched has lower profit and lower welfare than some strictly matched menu.
In particular, every feasible menu that maximizes profit or welfare satisfies strict matching.
\end{proposition}
The proof of \Cref{prop:strict_matching} uses the menu construction developed in Section~\ref{sec:properties} and appears in \Cref{apx:strict_matching}.
Intuitively, by adjusting only the wages, the platform can preserve consumer choices while reducing supply to match demand at every quality.
Whenever supply exceeds demand, this strictly lowers production costs and provider compensation without reducing consumer payments, improving both welfare and profit.
From now on, we restrict attention to strictly matched menus.

\section{Characterization of Optimal Menus}
\label{sec:properties}
This section characterizes optimal menus.
We first show that every trading distribution can be implemented by a strictly matched menu that is optimal for both profit and social welfare among the menus implementing it (\Cref{lem:implementation}), and we express the resulting profit and social welfare as integrals over quality (\Cref{thm:characterization}).
Menu design thus becomes an optimization over trading distributions.
We then show that replacing the technological cost function by its greatest convex minorant, which is determined by the technology alone, preserves both optimal profit and optimal social welfare, and that each objective admits an optimal menu that posts only qualities at which the two costs coincide (\Cref{thm:commonconvexification}).
Finally, we give a necessary and sufficient condition for finite optimal menus (\Cref{prop:finite_optimal_menu}) and show that some instances require infinitely many qualities (\Cref{ex:infinite_optimal_menu}).

\subsection{Optimal menus implementing a trading distribution}
\label{sec:reduction}
Every strictly matched menu induces a trading distribution, so menu design can proceed in two steps: choose a trading distribution, and then find the most profitable menu that induces it.
This subsection solves the second step for an arbitrary trading distribution, and the next subsection solves the first.

\begin{definition}[Implementation]\label{def:implementation}
Let $\var:(0,\qualityU]\to[0,1]$ be nonincreasing and left-continuous.
A menu $\menu$ \emph{implements} $\var$ if $\menu$ satisfies strict matching and the induced trading distribution $\tradedist=\var$.
\end{definition}

A menu implementing a given trading distribution $\var$ can be constructed as follows.
Assign higher qualities to consumers with higher values and to providers with lower costs.
Thus the top mass $\var(\quality)$ of consumers trades at quality at least $\quality$, and the consumer who chooses quality $\quality$ has value type $\inversevaldis(1-\var(\quality))$.
A consumer of type $\val$ maximizes $\val\cdot\quality-\price(\quality)$, so the first-order condition $\price'(\quality)=\val$ holds at her chosen quality; substituting $\val=\inversevaldis(1-\var(\quality))$ and integrating from $\price(0)=0$ gives the price schedule
\begin{align}
\price(\quality)\deq\int_0^{\quality}\inversevaldis(1-\var(\qualityvar))\,\dd\qualityvar.\label{eq:price}
\end{align}
Similarly, the providers' first-order condition $\wage'(\quality)=\costpar\cdot\effort'(\quality)$ gives the wage schedule
\begin{align}
\wage(\quality)\deq\int_0^{\quality}\effort'(\qualityvar)\cdot\inversecostdis(\var(\qualityvar))\,\dd\qualityvar.\label{eq:compensation}
\end{align}
Together, \cref{eq:price,eq:compensation} define the menu $(\qualities,\price,\wage)$, which posts every quality $\quality\in\qualities$.

\Cref{lem:implementation} below verifies that this menu indeed implements $\var$, and moreover, shows that this menu is optimal with respect to social welfare and profit among all menus implementing $\var$.
Intuitively, every menu implementing $\var$ assigns the same quality to almost every type, so all such menus yield the same social welfare and differ only in their transfers.
The menu in \cref{eq:price,eq:compensation} leaves each trader the least utility needed to keep her choice.
Thus the platform profit, which equals the social welfare minus the traders' utilities, is the largest. 
The proof of \Cref{lem:implementation} appears in \Cref{apx:implementation}.

\begin{theorem}\label{lem:implementation}
For an instance $(\qualities,\effort,\valdis,\costdis)$ and any nonincreasing left-continuous function $\var:(0,\qualityU]\to[0,1]$, the menu $(\qualities,\price,\wage)$ with the transfers in \cref{eq:price,eq:compensation} is strictly matched and implements $\var$.
Any other menu implementing $\var$ has the same social welfare as $(\qualities,\price,\wage)$, and $(\qualities,\price,\wage)$ maximizes platform profit among all menus implementing $\var$.
\end{theorem}

Thus, by \Cref{lem:implementation}, the platform can shape the market into any trading distribution $\var$ by designing a strictly matched menu defined by \cref{eq:price,eq:compensation}, and this implementation is optimal for both profit and social welfare among all menus implementing $\var$.
Next, we express the resulting platform profit and social welfare directly in terms of the trading distribution.

\begin{theorem}\label{thm:characterization}
For an instance $(\qualities,\effort,\valdis,\costdis)$ and every nonincreasing left-continuous function $\var:(0,\qualityU]\to[0,1]$, the optimal platform profit $\profit(\var)$ and the optimal social welfare $\welfare(\var)$ among menus implementing $\var$ are
\begin{align}
\profit(\var)&=\int_0^{\qualityU}\var(\quality)\cdot\left[\inversevaldis(1-\var(\quality))-\effort'(\quality)\cdot\inversecostdis(\var(\quality))\right]\,\dd\quality,\label{eq:profit-reduced}\\
\welfare(\var)&=\int_0^{\qualityU}\int_0^{\var(\quality)}\left[\inversevaldis(1-\massvar)-\effort'(\quality)\cdot\inversecostdis(\massvar)\right]\,\dd\massvar\,\dd\quality.\label{eq:welfare-reduced}
\end{align}
\end{theorem}

\Cref{thm:characterization} converts optimal menu design into an optimization over trading distributions.
Its proof in \Cref{apx:characterization} decomposes consumer payments, provider compensations, and production costs into quality increments and swaps the integration order.

\subsection{Optimal trading distributions via cost convexification}
By \Cref{lem:implementation,thm:characterization}, maximizing platform profit or social welfare over strictly matched menus is equivalent to an optimization over functions:
\begin{align}
\begin{aligned}
&\max_{\var:(0,\qualityU]\to[0,1]}\profit(\var)\quad\text{or}\quad\max_{\var:(0,\qualityU]\to[0,1]}\welfare(\var)\\
&\text{s.t.\ $\var$ is nonincreasing and left-continuous}.
\end{aligned}\label{eq:functional-optimization}
\end{align}
To solve this functional optimization problem, observe that at each quality $\quality$, the integrands of \cref{eq:profit-reduced,eq:welfare-reduced} depend on $\quality$ only through two numbers: the trading mass $\var(\quality)$ and the marginal technological cost $\effort'(\quality)$.
For a trading mass $\testmass\in[0,1]$ and a marginal technological cost $\slope\geq0$, define the \emph{profit} and \emph{welfare contributions} as the integrands with $\var(\quality)$ and $\effort'(\quality)$ replaced by $\testmass$ and $\slope$:
\begin{align*}
\profitctrib(\testmass;\slope)&\deq\testmass\cdot\left[\inversevaldis(1-\testmass)-\slope\cdot\inversecostdis(\testmass)\right],\\
\welfarectrib(\testmass;\slope)&\deq\int_0^{\testmass}\left[\inversevaldis(1-\massvar)-\slope\cdot\inversecostdis(\massvar)\right]\,\dd\massvar.
\end{align*}
Hence, the two objectives in \Cref{thm:characterization} take the compact forms
\begin{align*}
\profit(\var)=\int_0^{\qualityU}\profitctrib(\var(\quality);\effort'(\quality))\,\dd\quality,
\quad
\welfare(\var)=\int_0^{\qualityU}\welfarectrib(\var(\quality);\effort'(\quality))\,\dd\quality.
\end{align*}
The next theorem shows that the functional optimization in \cref{eq:functional-optimization} decouples across qualities once the technological cost is convexified: maximizing each contribution over the trading mass separately at every quality yields an optimal trading distribution for profit or welfare, respectively.

\begin{definition}[Convexified technological cost]\label{def:convexified_cost}
The \emph{convexified technological cost function} $\convexeffort$ is the greatest convex minorant of the technological cost function $\effort$, that is, the greatest convex function on $\qualities$ that lies below $\effort$.
A quality $\quality\in\qualities$ is a \emph{contact quality} if $\convexeffort(\quality)=\effort(\quality)$.
\end{definition}
On every open interval where the two costs differ, the convexified cost is affine and the interval endpoints are contact qualities, as illustrated in \Cref{fig:common_convexification}.
The derivative $\convexslope(\quality)$ exists at all but countably many qualities $\quality$ and is positive and nondecreasing.


\begin{theorem}\label{thm:commonconvexification}
For an instance $(\qualities,\effort,\valdis,\costdis)$, optimal trading distributions $\profitopt$ for profit and $\welfareopt$ for social welfare are obtained as:
\begin{align*}
\profitopt(\quality)&=\max\argmax_{\testmass\in[0,1]}\profitctrib(\testmass;\convexslope(\quality)),\\
\welfareopt(\quality)&=\max\argmax_{\testmass\in[0,1]}\welfarectrib(\testmass;\convexslope(\quality)),
\end{align*}
for almost every $\quality\in(0,\qualityU)$.
Moreover, for each objective, there exists an optimal menu with posted quality set $\posted\subseteq\{\quality\in\qualities:\effort(\quality)=\convexeffort(\quality)\}$.
\end{theorem}

Economically, \Cref{thm:commonconvexification} shows that optimal menus skip every interval of qualities on which $\effort$ exceeds $\convexeffort$: there, $\convexslope$ is constant, so both optimal trading distributions are constant and no trade occurs.
Thus, when the technological cost is strictly concave on an interval, that is, when the marginal technological cost decreases over this interval, the platform need not post any quality in it other than its endpoints, whatever the objective and type distributions.

The proof of \Cref{thm:commonconvexification} replaces technological cost by its greatest convex minorant and optimizes the scalar contributions along the resulting nondecreasing marginal technological costs.
Higher marginal technological costs favor smaller trading masses, so these independently optimal masses satisfy the required ordering.
The main issue is to return to the original technology without changing either objective: constancy of the trading mass on noncontact intervals supplies this equality, and \Cref{lem:implementation} supplies bounded transfers.
This argument parallels the ironing technique of \citet{M-81}, applied along qualities instead of types: the marginal technological cost $\effort'$ plays the role of the virtual value, and convexifying $\effort$ corresponds to convexifying the integral of the virtual value.
Unlike that ironing, which depends on the value distribution, cost convexification depends only on the technology.
The complete proof appears in \Cref{apx_commonconvexification}.

\subsubsection{Finite optimal menus}
In a continuous quality space, \Cref{thm:commonconvexification} does not by itself imply that an unrestricted optimum uses finitely many qualities.
A trading distribution is \emph{finitely supported} if trade occurs at only finitely many positive qualities, which we call its \emph{active qualities}.
It is then constant between consecutive active qualities, its jumps are the masses trading at those qualities, and it is zero above the highest one.
Unused offers can be deleted, so existence of a finite optimal posted menu is equivalent to existence of a finitely supported optimal trading distribution.

\begin{proposition}\label{prop:finite_optimal_menu}
For an instance $(\qualities,\effort,\valdis,\costdis)$, a finite menu attains unrestricted optimal profit if and only if there exists a finitely supported trading distribution whose active qualities are contacts and mass maximizes the scalar profit contribution at almost every quality:
\begin{align}
\var(\quality)\in\argmax_{\testmass\in[0,1]}\profitctrib(\testmass;\convexslope(\quality))
\quad\text{for almost every }\quality\in(0,\qualityU).\label{eq:finite_profit_criterion}
\end{align}
The corresponding criterion for unrestricted welfare replaces the profit contribution in \cref{eq:finite_profit_criterion} by the welfare contribution.
\end{proposition}

The proof in \Cref{apx:finite_menus} derives this criterion directly from the two equality conditions in the proof of \Cref{thm:commonconvexification}.
The criterion concerns existence of a finite optimal selection.
Finiteness of the contact set is sufficient but not necessary: infinitely many qualities can be technologically undominated while only finitely many are used optimally.
What matters is whether scalar maximizing masses can be chosen to form a finitely supported trading distribution.

For example, suppose the convexified cost is piecewise linear with $\convextiers$ pieces, as it is whenever the original cost is piecewise linear with finitely many pieces.
Then the marginal technological cost is constant on each piece, so both objectives admit optimal menus with at most $\convextiers$ positive qualities, located at the right endpoints of the pieces.
The following example shows that continuous strictly convex cost can instead require an optimal trading distribution with infinite support.

\begin{example}[Quadratic technological cost]\label{ex:infinite_optimal_menu}
Let $\qualityU=1$ and $\effort(\quality)=K\quality^2$ for $K>0$, with marginal technological cost $\effort'(\quality)=2K\quality$, and let both types be uniform on $[0,1]$.
The profit contribution $\profitctrib(\testmass;\slope)=\testmass-(1+\slope)\cdot\testmass^2$ has the unique maximizer $\testmass=1/(2\cdot(1+\slope))$; at $\slope=\effort'(\quality)=2K\quality$, the optimal trading distribution is $\profitopt(\quality)=1/(2(1+2K\quality))$.
It is strictly decreasing, so trade occurs at infinitely many qualities and no finite menu is optimal; welfare behaves the same way, with $\welfareopt(\quality)=1/(1+2K\quality)$.
\end{example}

Strict convexity of technological cost alone is not a general criterion for infinite optimal support: the scalar maximizing masses may be constant, including zero trade.
\Cref{prop:finite_optimal_menu} states the exact condition, while piecewise linearity of the convexified cost is a sufficient condition that depends only on the cost geometry.
The following example shows that the exact condition can hold even when the sufficient condition fails.

\begin{example}[Infinitely many contacts and a finite optimum]\label{ex:finite_despite_contacts}
Keep the cost and the provider distribution of \Cref{ex:infinite_optimal_menu} and set $K=1$, so every quality is still a contact, but let consumer values be uniform on $[6,7]$.
Now $\profitctrib(\testmass;\slope)=7\cdot\testmass-(1+\slope)\cdot\testmass^2$ is increasing on $[0,1]$ for every marginal technological cost $\slope\leq2$, so its maximizer is $\testmass=1$, and the same holds for welfare.
Hence all traders trade at quality one through the single offer $\price(1)=6$, $\wage(1)=1$: a strictly convex cost can still have a finite optimal menu.
\end{example}

\section{Efficiency of Profit Maximization}
\label{sec:efficiency}
In this section we study the efficiency loss caused by profit maximization.
Let $\profitopt$ and $\welfareopt$ be the selected profit- and welfare-optimal trading distributions in \Cref{thm:commonconvexification}.
When $\welfare(\welfareopt)>0$, define the welfare ratio $\welfareratio
    \deq\frac{\welfare(\profitopt)}{\welfare(\welfareopt)}$.
    
We first show that the welfare benchmark requires no platform subsidy.
We then prove that \emph{regularity} alone provides no positive uniform lower bound on $\welfareratio$, whereas \emph{MHR} guarantees a tight bound $\welfareratio\geq1/(\ee-1)$, where $\ee$ is the base of the natural logarithm.


\begin{proposition}
\label{prop:weak_budget_balance}
For an instance $(\qualities,\effort,\valdis,\costdis)$, the selected welfare-optimal trading distribution $\welfareopt$ can be implemented under the original technological cost with nonnegative platform profit: $\profit(\welfareopt)\geq 0$.
\end{proposition}

The efficiency loss measured by $\welfareratio$ therefore arises from maximizing profit, rather than the requirement of nonnegative profit.
The proof appears in \Cref{apx:wbb}.

\begin{definition}[Regularity]
    A consumer's value distribution $\valdis$ is \emph{regular} if the consumer revenue curve $\testmass\cdot\inversevaldis(1-\testmass)$ is concave.
    A provider's cost distribution $\costdis$ is \emph{regular} if the provider expenditure curve $\testmass\cdot\inversecostdis(\testmass)$ is convex.
\end{definition}

\begin{proposition}\label{prop:regular_insufficient}
    For any $\approxrate>0$, there exists an instance $(\qualities,\effort,\valdis,\costdis)$ with both $\valdis$ and $\costdis$ regular and $\welfare(\welfareopt)>0$, such that the welfare ratio $\welfareratio<\approxrate$.
\end{proposition}

Regularity controls the shape of the profit objective but does not bound the surplus forgone by the platform.
The following example illustrates how the welfare ratio can approach zero. 
The proof appears in \Cref{apx:regular_insufficient}.

\begin{example}\label{example:regular_counterexample}
    Let $\qualityU=1$ and $\effort(\quality)=\quality$ at every feasible quality $\quality\in[0,1]$.
    Thus the marginal technological cost is $\slope=1$.
    Fix a parameter $0<\parameter<1/2$.
    Define the consumer and provider distributions through  quantile functions as $\inversecostdis(\testmass)=\testmass$ for $\testmass\in[0,1]$ and
    \begin{align*}
        \inversevaldis(1-\testmass)
        &\deq
        \begin{dcases}
            \frac{1}{\parameter}-\testmass,
            &0\leq\testmass\leq\parameter,\\
            (1+\parameter)\cdot\frac{1-\testmass}{\testmass},
            &\parameter<\testmass\leq1.
        \end{dcases}
    \end{align*}
    The support of $\valdis$ is $[0,1/\parameter]$ and the support of $\costdis$ is $[0,1]$.
    Both distributions are absolutely continuous and strictly increasing.
\end{example}

\subsection{A constant guarantee under MHR}
The negative result under regularity motivates the stronger MHR conditions.
Our results show that under MHR assumptions, profit maximization preserves a constant fraction of the optimal social welfare and the guarantee is tight.

\begin{definition}[Monotone hazard rate]
    A consumer's value distribution $\valdis$ satisfies \emph{monotone hazard rate (MHR)} if its hazard rate $\valhrate_{\valdis}(\val)\deq\valdens(\val)/(1-\valdis(\val))$ is nondecreasing.
    A provider's cost distribution $\costdis$ satisfies MHR if its reverse hazard rate $\costhrate_{\costdis}(\costpar)\deq\costdens(\costpar)/\costdis(\costpar)$ is nonincreasing.
\end{definition}

\begin{theorem}\label{thm:mhr_ratio}
    For an instance $(\qualities,\effort,\valdis,\costdis)$, if both $\valdis$ and $\costdis$ satisfy MHR and $\welfare(\welfareopt)>0$, the welfare ratio $\welfareratio\geq1/(\ee-1)$.
    
    The ratio is tight: there exist a sequence of instances whose ratios converge to $1/(\ee-1)$.
\end{theorem}

Intuitively, the MHR condition limits how slowly the gap between the marginal consumer's value and the marginal provider's cost can decline as trade expands, preventing the excluded trades from accounting for an arbitrarily large share of the gains from trade.
The formal proof uses the following technical lemma and can be found in \Cref{apx:mhr_ratio}.

\begin{lemma}\label{thm:profit_vs_welfare}
    For an instance $(\qualities,\effort,\valdis,\costdis)$, the selected optimal trading distributions satisfy
    \begin{align*}
        \profitopt(\quality)\leq\welfareopt(\quality),
        \qquad 0<\quality\leq\qualityU.
    \end{align*}
\end{lemma}

This comparison holds under the baseline assumptions, without regularity or MHR. 
The proof appears in \Cref{apx_profit_vs_welfare}.

\section{Limited Menus and Computation}
\label{sec:limitedmenus}
The characterization in \Cref{sec:properties} does not directly provide a finite computational procedure, particularly when an optimal menu requires infinitely many qualities.
In practice, platforms typically offer finite menus that participants can readily compare and choose from.
We therefore study profit maximization over menus offering at most $\menusize\in\naturals_{+}$ positive qualities excluding the outside option.

Let $\oprofit$ be the unrestricted optimal profit and $\oprofit_{\menusize}$ the supremum under this restriction.
Because the proof of \Cref{prop:strict_matching} weakly improves every feasible menu with finitely many qualities by a strictly matched menu that posts the same qualities, $\oprofit_{\menusize}$ is the same over feasible and strictly matched menus.
We establish a tight $\menusize/\tiers$ profit guarantee when the smallest unrestricted optimal menu has $\tiers\in\naturals_{+}$ positive qualities, and give a dynamic program with an additive profit guarantee for general instances.

\subsection{A tight guarantee for finite optimal menus}

We first quantify the profit loss from menu size restriction when the unrestricted optimal menu is finite.

\begin{theorem}\label{thm:restrictmenu_profitloss}
    For an instance $(\qualities,\effort,\valdis,\costdis)$, if the smallest unrestricted optimal menu has $\tiers\in\naturals_{+}$ positive qualities, for every integer $1\leq\menusize\leq\tiers$, $\oprofit_{\menusize}\geq\menusize/\tiers \cdot\oprofit$.
    
    The ratio is tight: for any $1\leq\menusize\leq\tiers$, there exist a sequence of instances whose ratios converge to $\menusize/\tiers$.
\end{theorem}

The proof constructs a feasible $\menusize$-quality menu by successively removing qualities with small incremental profit contributions from a finite unrestricted optimum.
Economically, the result quantifies the worst-case cost of limiting menu size, even when the platform optimally chooses which qualities to offer.
The proof and tightness construction appear in \Cref{app:limitedmenus_continuous}.

\subsection{Additive approximation and computation}

The multiplicative guarantee in \Cref{thm:restrictmenu_profitloss} relies on the existence of a finite unrestricted optimal menu.
Without this assumption, \Cref{ex:infinite_optimal_menu} shows that no positive multiplicative guarantee depending only on $\menusize$ holds uniformly across instances.\footnote{In \Cref{ex:infinite_optimal_menu}, each constant-mass interval $(a,b]$ contributes at most $(b-a)/[4(1+K(a+b))]\leq1/(4K)$.
Thus $\oprofit_{\menusize}/\oprofit\leq2\menusize/\log(1+2K)\to0$ as $K\to\infty$ for every fixed $\menusize$.}
We therefore develop a dynamic program that offers at most $\menusize$ qualities with a bounded additive profit loss relative to the unrestricted optimum.

Write $\consumermean$ for the mean consumer value.
For an accuracy parameter $\approxrate>0$, let $\menu_{\menusize,\approxrate}=(\posted_{\menusize,\approxrate},\price_{\menusize,\approxrate},\wage_{\menusize,\approxrate})$ denote the output menu and $\profit(\menu_{\menusize,\approxrate})$ its profit.

We assume exact real arithmetic and access to technological cost and quantile evaluations, plus the scalar profit oracle defined below.
Given any marginal technological cost $\slope$, the oracle returns
\begin{align*}
    \oraclevalue(\slope)
    \deq\max_{\testmass\in[0,1]}\profitctrib(\testmass;\slope), \quad
    \oracle(\slope)\deq\max\argmax_{\testmass\in[0,1]}\profitctrib(\testmass;\slope).
\end{align*}

\begin{theorem}
    \label{thm:compute_restricted_menu}
    \label{thm:continuous_additive_approximation}
    \label{cor:menu_size_additive}
    For an instance $(\qualities,\effort,\valdis,\costdis)$, any integer $\menusize\geq1$ and accuracy parameter $\approxrate>0$, under the scalar profit oracle model, there is an algorithm that computes a strictly matched menu $\menu_{\menusize,\approxrate}$ with at most $\menusize$ positive qualities such that
    \begin{align*}
        \profit(\menu_{\menusize,\approxrate})
        \geq\oprofit-\frac{\qualityU\cdot\consumermean}{\menusize}-\approxrate.
    \end{align*}
    If the consumer distribution $\valdis$ and the provider distribution $\costdis$ additionally satisfy MHR, the same output menu has social welfare
    \begin{align*}
        \welfare(\menu_{\menusize,\approxrate})
        \geq\frac{\welfare(\welfareopt)}{\ee}
        -\frac{\qualityU\cdot\consumermean}{\menusize}-\approxrate.
    \end{align*}
    The algorithm uses $\bO{\approxstage^2}$ oracle calls and $\bO{\min\{\menusize,\approxstage\}\approxstage^2}$ arithmetic operations, where $\approxstage\deq\max\left\{1,
            \left\lceil\qualityU\cdot\valU/\approxrate\right\rceil\right\}$
\end{theorem}

Overall, the algorithm first discretizes the quality space, then convexifies the grid cost and retains its contact qualities.
Finally, a dynamic program groups consecutive quality increments to select at most $\menusize$ positive qualities and recover their prices and wages.

Below we show the full algorithm and a useful technical lemma used for welfare approximation in MHR condition while the proof of \Cref{thm:compute_restricted_menu} appears in \Cref{app:restricted_dp_continuous}.

\xhdr{Algorithm.}
Given $\menusize$ and $\approxrate$, the algorithm proceeds in three steps. 

\xhdr{Step 1: Quality discretization.}
Set $\discresize=\qualityU/\approxstage$ and discretize the quality interval into the uniform grid $\qualities_{\approxstage}=\{0,\discresize,2\discresize,\ldots,\approxstage\discresize\}$.
We restrict candidate qualities to $\qualities_{\approxstage}$ while retaining the menu-size limit $\menusize$.
On this grid, replace $\effort$ by the piecewise-linear interpolation of $(\quality,\effort(\quality))$, $\quality\in\qualities_{\approxstage}$.

\xhdr{Step 2: Convexification.}
Compute the greatest convex minorant of this interpolated cost and retain its contact grid qualities.
Let the retained qualities be $0=\qTier{0}<\cdots<\qTier{\convextiers}=\qualityU$.
By convexity, slopes between consecutive retained qualities are nondecreasing.

\xhdr{Step 3: Dynamic programming and menu recovery.}
For every $0\leq\leftindex<\rightindex\leq\convextiers$, define
\begin{align*}
    \slope_{\leftindex,\rightindex}
    \deq\frac{\effort(\qTier{\rightindex})-\effort(\qTier{\leftindex})}
        {\qTier{\rightindex}-\qTier{\leftindex}}, \quad
    \profit(\leftindex,\rightindex)
    \deq(\qTier{\rightindex}-\qTier{\leftindex})
        \oraclevalue(\slope_{\leftindex,\rightindex}).
\end{align*}
$\profit(\leftindex,\rightindex)$ is the maximum profit contribution when trading mass is constant on $(\qTier{\leftindex},\qTier{\rightindex}]$. 
Let $\dpfunc[\rightindex,\groupcount]$ be the maximum total contribution from partitioning the first $\rightindex$ intervals into exactly $\groupcount$ nonempty consecutive groups. 
Setting $\dpfunc[\rightindex,1] = \profit(0,\rightindex)$, we compute
\begin{align*}
    \dpfunc[\rightindex,\groupcount]
    = \max_{\groupcount-1\leq\leftindex<\rightindex}
        \{\dpfunc[\leftindex,\groupcount-1]
            +\profit(\leftindex,\rightindex)\},
\end{align*}
for $2\leq\groupcount\leq\min\{\menusize,\rightindex\}$.
The optimal restricted grid profit is
\begin{align*}
\max_{\substack{1\leq\rightindex\leq\convextiers\\
            1\leq\groupcount\leq\min\{\menusize,\rightindex\}}}\left\{0,
        \dpfunc[\rightindex,\groupcount]\right\}.
\end{align*}
Backtracking the dynamic program yields at most $
\menusize$ qualities, with trading masses determined by $\oracle(\slope_{\leftindex,\rightindex})$. 
The original menu is then recovered using \cref{eq:price,eq:compensation} with quantile evaluations.

\begin{lemma}\label{lem:mhr_profit_welfare} 
    For an instance $(\qualities,\effort,\valdis,\costdis)$,
    if $\valdis$ and $\costdis$ satisfy MHR, the unrestricted optimal profit $\oprofit$ satisfies
    \begin{align*}
        \oprofit\geq\frac{\welfare(\welfareopt)}{\ee}.
    \end{align*}
\end{lemma}

The above comparison converts a profit guarantee into a welfare guarantee under MHR.
The proof appears in \Cref{app:mhr_profit_welfare}.

\section{Conclusion and Discussion}
\label{sec:conclusion}
In this work, we study the two-sided screening problem in quality-differentiated markets.
We characterize optimal two-sided quality menus through trading distributions and a convexification of technological cost.
The characterization identifies when finite optimal menus exist and supports tight guarantees for welfare under profit maximization and profit under menu-size restrictions.
For general continuous-quality instances, our dynamic program provides an additive profit guarantee and, under MHR, a welfare guarantee for the same output menu.

Future work could study menu design when type distributions must be learned from transaction data, including how estimation errors affect profit, welfare, and market clearing.
Another direction is to allow waiting, rationing, or quality upgrades and examine whether the cost convexification and approximation guarantees extend to these settings.

\bibliography{mybibfile}

\appendix

\section{Omitted Proofs in Section \ref{sec:preliminary} and \ref{sec:properties}.}
\label{apx:properties}

\subsection[Implementation]{Proof of \Cref{lem:implementation}}\label{apx:implementation}

\begin{proof}[Proof of \Cref{lem:implementation}]
We first show that menus sort types, so that a trading distribution determines, up to null sets, which types trade at each quality threshold.
We then verify that the menu in \cref{eq:price,eq:compensation} implements the given function.
Finally, we show that every other menu implementing the function has the same welfare and a profit reduced by its endpoint utilities.

\xhdr{Sorting.}
Fix a menu and consider consumer types $\val_1<\val_2$ with maximized choices $\quality_1$ and $\quality_2$.
Optimality gives $\val_1\cdot\quality_1-\price(\quality_1)\geq\val_1\cdot\quality_2-\price(\quality_2)$ and $\val_2\cdot\quality_2-\price(\quality_2)\geq\val_2\cdot\quality_1-\price(\quality_1)$.
Adding yields $(\val_2-\val_1)\cdot(\quality_2-\quality_1)\geq0$, so $\quality_2\geq\quality_1$.
For provider types $\costpar_1<\costpar_2$ with maximizing choices $\quality_1$ and $\quality_2$, adding the optimality inequalities $\wage(\quality_1)-\costpar_1\cdot\effort(\quality_1)\geq\wage(\quality_2)-\costpar_1\cdot\effort(\quality_2)$ and $\wage(\quality_2)-\costpar_2\cdot\effort(\quality_2)\geq\wage(\quality_1)-\costpar_2\cdot\effort(\quality_1)$ yields $(\costpar_2-\costpar_1)\cdot[\effort(\quality_2)-\effort(\quality_1)]\leq0$, so $\quality_2\leq\quality_1$ because $\effort$ is strictly increasing.
Hence, on the full-measure set of types with maximizing choices, the consumer choice map is nondecreasing and the provider choice map is nonincreasing.
Now let the menu be strictly matched with trading distribution $\var$, and fix a positive quality $\quality$.
Up to a null set, the consumers choosing quality at least $\quality$ form an upper interval of types with probability $\var(\quality)$; because $\valdis$ is continuous and strictly increasing, they are, up to a null set, the types with $1-\valdis(\val)\leq\var(\quality)$.
Up to a null set, the providers choosing quality at least $\quality$ form a lower interval of types with probability $\var(\quality)$; because $\costdis$ is continuous and strictly increasing, they are, up to a null set, the types with $\costdis(\costpar)\leq\var(\quality)$.

\xhdr{Implementation.}
Fix a nonincreasing left-continuous function $\var:(0,\qualityU]\to[0,1]$, and post every quality in $\qualities$ with the transfers in \cref{eq:price,eq:compensation}.
These transfers are continuous in quality and satisfy $0\leq\price(\quality)\leq\valU\cdot\qualityU$ and $0\leq\wage(\quality)\leq\costU\cdot\effort(\qualityU)$, because both quantile functions take values in bounded nonnegative supports.
Define the choice maps
\begin{align*}
\qC(\val)&\deq\sup\bigl(\{0\}\cup\{\qualityvar\in(0,\qualityU]:\var(\qualityvar)\geq1-\valdis(\val)\}\bigr),\\
\qP(\costpar)&\deq\sup\bigl(\{0\}\cup\{\qualityvar\in(0,\qualityU]:\var(\qualityvar)\geq\costdis(\costpar)\}\bigr).
\end{align*}
The consumer map is nondecreasing and the provider map is nonincreasing, so both are measurable.
Because $\var$ is nonincreasing, each set in these definitions contains every positive quality below any of its elements; because $\var$ is also left-continuous, each nonempty set contains its supremum.
Hence, for every positive quality $\quality$,
\begin{align}
\qC(\val)\geq\quality&\Longleftrightarrow\var(\quality)\geq1-\valdis(\val),\label{eq:consumer_threshold}\\
\qP(\costpar)\geq\quality&\Longleftrightarrow\var(\quality)\geq\costdis(\costpar).\label{eq:provider_threshold}
\end{align}
For every consumer type $\val$, provider type $\costpar$, and quality $\quality$, these transfers give
\begin{align*}
\val\cdot\quality-\price(\quality)&=\int_0^{\quality}\left[\val-\inversevaldis(1-\var(\qualityvar))\right]\,\dd\qualityvar,\\
\wage(\quality)-\costpar\cdot\effort(\quality)&=\int_0^{\quality}\left[\inversecostdis(\var(\qualityvar))-\costpar\right]\,\dd\effort(\qualityvar).
\end{align*}
By \cref{eq:consumer_threshold}, the first integrand is nonnegative at qualities $\qualityvar\leq\qC(\val)$, because there $\inversevaldis(1-\var(\qualityvar))\leq\val$, and negative at qualities $\qualityvar>\qC(\val)$.
The consumer's utility is therefore nondecreasing in quality up to $\qC(\val)$ and nonincreasing beyond it, so $\qC(\val)$ maximizes it.
By \cref{eq:provider_threshold}, the second integrand is nonnegative at qualities $\qualityvar\leq\qP(\costpar)$, because there $\inversecostdis(\var(\qualityvar))\geq\costpar$, and negative at qualities $\qualityvar>\qP(\costpar)$.
Since $\effort$ is strictly increasing, the provider's utility is therefore nondecreasing in quality up to $\qP(\costpar)$ and nonincreasing beyond it, so $\qP(\costpar)$ maximizes it.
Thus every type has a maximizing choice under the menu $(\qualities,\price,\wage)$, the maps above are measurable maximizing selections, and expected transfers are finite.
Since $\valdis$ and $\costdis$ are continuous, both $1-\valdis(\val)$ and $\costdis(\costpar)$ are uniformly distributed on $[0,1]$.
By \cref{eq:consumer_threshold,eq:provider_threshold}, a consumer and a provider each choose quality at least $\quality$ with probability exactly $\var(\quality)$, for every positive quality $\quality$.
The demand and supply distributions therefore both equal $\var$, which establishes strict matching and implementation.

\xhdr{Comparison with other menus.}
Consider any menu $\menu'$ implementing $\var$, with consumer and provider indirect utilities $\Ucons$ and $\Uprov$.
Applying the sorting step at every rational quality threshold and comparing with \cref{eq:consumer_threshold,eq:provider_threshold} shows that its choice maps agree with the maps $\qC$ and $\qP$ of the implementation step at almost every type.
Consequently, its social welfare equals that of $(\qualities,\price,\wage)$, since by \cref{eq:welfare} welfare depends only on the qualities assigned to types.
The indirect utility $\Ucons$ is a supremum of affine functions of the consumer type with slopes in $[0,\qualityU]$, and it is finite because almost every type has a maximizing choice.
Hence $\Ucons$ is convex and Lipschitz, and at almost every type its derivative equals the menu's maximizing choice, which is $\qC(\val)$, because a maximizing choice is a subgradient.
Likewise, $\Uprov$ is a finite supremum of affine functions of the provider type with slopes in $[-\effort(\qualityU),0]$, so it is convex and Lipschitz with derivative $-\effort(\qP(\costpar))$ at almost every type.
Writing $\typevar$ for a type integration variable, absolute continuity yields the envelope identities
\begin{align*}
\Ucons(\val)&=\Ucons(\valL)+\int_{\valL}^{\val}\qC(\typevar)\,\dd\typevar,\\
\Uprov(\costpar)&=\Uprov(\costU)+\int_{\costpar}^{\costU}\effort(\qP(\typevar))\,\dd\typevar.
\end{align*}
Taking $\val=\valL$ and $\costpar=\costU$ in the consumer and provider utility formulas of the implementation step, both integrands are nonpositive because $\inversevaldis\geq\valL$ and $\inversecostdis\leq\costU$.
Since the outside option yields zero, the menu $(\qualities,\price,\wage)$ gives zero utility to the lowest consumer type and to the highest provider type.
The envelope identities, which hold for both menus with the same choice maps, therefore show that the alternative menu's consumer indirect utility exceeds that under $(\qualities,\price,\wage)$ by exactly $\Ucons(\valL)$ at almost every consumer type, and its provider indirect utility exceeds that under $(\qualities,\price,\wage)$ by exactly $\Uprov(\costU)$ at almost every provider type.
Because both menus assign the same qualities at almost every type, the alternative menu collects $\Ucons(\valL)$ less from almost every consumer and pays $\Uprov(\costU)$ more to almost every provider.
With unit masses, its profit equals the profit of $(\qualities,\price,\wage)$ minus $\Ucons(\valL)+\Uprov(\costU)$.
The outside option gives $\Ucons(\valL)\geq0$ and $\Uprov(\costU)\geq0$, so no menu implementing $\var$ earns more profit than $(\qualities,\price,\wage)$.
\end{proof}

\subsection[Characterization]{Proof of \Cref{thm:characterization}}\label{apx:characterization}

\begin{proof}[Proof of \Cref{thm:characterization}]
Fix $\var$ and consider the menu $(\qualities,\price,\wage)$ in \cref{eq:price,eq:compensation}.
By \Cref{lem:implementation}, this menu implements $\var$, maximizes platform profit among all menus implementing $\var$, and has the same social welfare as every other such menu.
It therefore suffices to show that its profit and welfare equal the right-hand sides of \cref{eq:profit-reduced,eq:welfare-reduced}.
We use the choice maps $\qC$ and $\qP$ constructed in the proof of \Cref{lem:implementation} in \Cref{apx:implementation}.
By \cref{eq:consumer_threshold,eq:provider_threshold}, a consumer and a provider each choose quality at least $\quality$ with probability $\var(\quality)$, for every quality $\quality$.

\xhdr{Profit.}
By \cref{eq:profit}, the profit of this menu is its expected consumer payment minus its expected provider compensation.
The expected consumer payment is
\begin{align*}
\expect[\val]{\price(\qC(\val))}
&\overset{(a)}{=}\expect[\val]{\int_0^{\qC(\val)}\inversevaldis(1-\var(\quality))\,\dd\quality}\\
&\overset{(b)}{=}\expect[\val]{\int_0^{\qualityU}\inversevaldis(1-\var(\quality))\cdot\indicator{\qC(\val)\geq\quality}\,\dd\quality}\\
&\overset{(c)}{=}\int_0^{\qualityU}\expect[\val]{\inversevaldis(1-\var(\quality))\cdot\indicator{\qC(\val)\geq\quality}}\,\dd\quality\\
&\overset{(d)}{=}\int_0^{\qualityU}\inversevaldis(1-\var(\quality))\cdot\prob[\val]{\qC(\val)\geq\quality}\,\dd\quality\\
&\overset{(e)}{=}\int_0^{\qualityU}\var(\quality)\cdot\inversevaldis(1-\var(\quality))\,\dd\quality.
\end{align*}
Equality~(a) substitutes the price in \cref{eq:price}.
Equality~(b) extends the integral to the full quality interval, with the indicator removing all increments above the consumer's chosen quality.
Equality~(c) exchanges the order of expectation and integration.
Equality~(d) factors out the quantile value, which does not depend on the consumer type, and uses that the expected indicator equals the probability of its event.
By \Cref{def:demand_supply,def:strict_matching}, $\prob[\val]{\qC(\val)\geq\quality}=\demdist(\quality)=\var(\quality)$, proving equality~(e).
By a similar argument, using the compensation in \cref{eq:compensation}, the expected provider compensation is
\begin{align*}
\expect[\costpar]{\wage(\qP(\costpar))}=\int_0^{\qualityU}\var(\quality)\cdot\effort'(\quality)\cdot\inversecostdis(\var(\quality))\,\dd\quality.
\end{align*}
Subtracting the expected compensation from the expected payment gives \cref{eq:profit-reduced}.

\xhdr{Welfare.}
By \cref{eq:welfare}, the welfare of this menu is its expected consumer gross value minus its expected provider production cost.
Since $\qC(\val)=\int_0^{\qualityU}\indicator{\qC(\val)\geq\quality}\,\dd\quality$, the expected consumer gross value is
\begin{align*}
\expect[\val]{\val\cdot\qC(\val)}
&\overset{(a)}{=}\int_0^{\qualityU}\expect[\val]{\val\cdot\indicator{\qC(\val)\geq\quality}}\,\dd\quality\\
&\overset{(b)}{=}\int_0^{\qualityU}\int_{\inversevaldis(1-\var(\quality))}^{\valU}\val\,\dd\valdis(\val)\,\dd\quality\\
&\overset{(c)}{=}\int_0^{\qualityU}\int_0^{\var(\quality)}\inversevaldis(1-\massvar)\,\dd\massvar\,\dd\quality.
\end{align*}
Equality~(a) exchanges the order of expectation and integration.
Equality~(b) uses $\qC(\val)\geq\quality\Longleftrightarrow1-\valdis(\val)\leq\var(\quality)\Longleftrightarrow\val\geq\inversevaldis(1-\var(\quality))$ from \cref{eq:consumer_threshold}.
Equality~(c) changes to the mass variable $\massvar\deq1-\valdis(\val)$: $\val=\inversevaldis(1-\massvar)$ and $\dd\valdis(\val)=-\dd\massvar$.
The type limits become mass limits $\var(\quality),0$ and are reversed.
Since $\effort(0)=0$, we have $\effort(\qP(\costpar))=\int_0^{\qualityU}\indicator{\qP(\costpar)\geq\quality}\,\dd\effort(\quality)$, and the expected provider production cost is
\begin{align*}
\expect[\costpar]{\costpar\cdot\effort(\qP(\costpar))}
&\overset{(a)}{=}\int_0^{\qualityU}\effort'(\quality)\cdot\expect[\costpar]{\costpar\cdot\indicator{\qP(\costpar)\geq\quality}}\,\dd\quality\\
&\overset{(b)}{=}\int_0^{\qualityU}\effort'(\quality)\cdot\int_{\costL}^{\inversecostdis(\var(\quality))}\costpar\,\dd\costdis(\costpar)\,\dd\quality\\
&\overset{(c)}{=}\int_0^{\qualityU}\int_0^{\var(\quality)}\effort'(\quality)\cdot\inversecostdis(\massvar)\,\dd\massvar\,\dd\quality.
\end{align*}
Equality~(a) exchanges the order of expectation and integration and uses $\dd\effort(\quality)=\effort'(\quality)\,\dd\quality$.
Equality~(b) uses $\qP(\costpar)\geq\quality\Longleftrightarrow\costdis(\costpar)\leq\var(\quality)\Longleftrightarrow\costpar\leq\inversecostdis(\var(\quality))$ from \cref{eq:provider_threshold}.
Equality~(c) changes to the mass variable $\massvar\deq\costdis(\costpar)$: $\costpar=\inversecostdis(\massvar)$ and $\dd\costdis(\costpar)=\dd\massvar$, with mass limits $0,\var(\quality)$.
Subtracting the expected production cost from the expected gross value gives \cref{eq:welfare-reduced}.

\end{proof}
\subsection[Strict matching]{Proof of \Cref{prop:strict_matching}}\label{apx:strict_matching}

\begin{proof}[Proof of \Cref{prop:strict_matching}]
Let $\menu=(\posted,\price,\wage)$ be a feasible menu.
Since $\demdist(0)=\supdist(0)=1$ and feasibility gives $\demdist(\quality)\leq\supdist(\quality)$ at every positive quality $\quality$, if $\menu$ is not strictly matched, then $\demdist(\quality)<\supdist(\quality)$ at some positive quality $\quality$.
It therefore suffices to show that $\menu$ is weakly improved in both profit and welfare by a strictly matched menu, and strictly improved if $\demdist(\quality)<\supdist(\quality)$ at some positive quality $\quality$.
The second claim follows because every strictly matched menu is feasible, so a feasible menu that is not strictly matched maximizes neither profit nor welfare over feasible menus.
For use in \Cref{sec:limitedmenus}, we also show that this strictly matched menu can be chosen to post exactly $\posted$ when $\posted$ is finite.

\xhdr{Replacement.}
Let $\menu$ have choice maps $\qC$ and $\qP$ and indirect utilities $\Ucons$ and $\Uprov$.
The function $\demdist$ is nonincreasing and left-continuous on $(0,\qualityU]$, with values in $[0,1]$, because the event $\qC(\val)\geq\quality$ shrinks as $\quality$ increases and is the intersection of the corresponding events at all smaller positive qualities.
By \Cref{lem:implementation}, the menu defined by \cref{eq:price,eq:compensation} with $\demdist$ in place of $\var$ gives every type a maximizing choice and is strictly matched; we call it the replacement.
By the proof of \Cref{lem:implementation}, the choice maps of the replacement satisfy \cref{eq:consumer_threshold,eq:provider_threshold} with $\demdist$ in place of $\var$, and the replacement gives zero utility to the lowest consumer type and to the highest provider type.

\xhdr{Choices.}
By the sorting step in the proof of \Cref{lem:implementation}, the map $\qC$ is nondecreasing and the map $\qP$ is nonincreasing.
Fix a positive quality $\quality$.
Under $\menu$, the consumers who select a quality of at least $\quality$ therefore form an upper interval of types with probability $\demdist(\quality)$, and the providers who do so form a lower interval of types with probability $\supdist(\quality)\geq\demdist(\quality)$.
Because $\valdis$ and $\costdis$ are continuous and strictly increasing, \cref{eq:consumer_threshold,eq:provider_threshold} show that, except possibly at one boundary type on each side, the replacement selects a quality of at least $\quality$ at the same consumer types as $\menu$ and at a subset of the provider types at which $\menu$ does.
Since any two different qualities are separated by a positive rational quality, applying this at every positive rational quality shows that, except at countably many types, the replacement assigns the same quality as $\menu$ to every consumer type and a weakly lower quality to every provider type.

\xhdr{Comparison.}
Countable sets of types are null, both under the type distributions and for integrals over types.
The consumer half of the comparison step in the proof of \Cref{lem:implementation} uses only that the two menus assign the same quality to almost every consumer type.
Hence the replacement collects $\Ucons(\valL)\geq0$ more than $\menu$ from almost every consumer, while the expected gross value $\expect[\val]{\val\cdot\qC(\val)}$ is unchanged.
For providers, the envelope identity in that step gives, for either menu with its own wage schedule, provider choice map, and provider indirect utility,
\begin{align*}
\expect[\costpar]{\wage(\qP(\costpar))}=\expect[\costpar]{\costpar\cdot\effort(\qP(\costpar))}+\Uprov(\costU)+\expect[\costpar]{\int_{\costpar}^{\costU}\effort(\qP(\typevar))\,\dd\typevar}.
\end{align*}
For the replacement, the second term is zero and the provider choice map is weakly lower at almost every type; since $\effort$ is increasing, its first and third terms are therefore weakly lower than those of $\menu$.
The replacement thus pays weakly less to providers and incurs a weakly lower expected production cost $\expect[\costpar]{\costpar\cdot\effort(\qP(\costpar))}$.
By \cref{eq:profit,eq:welfare}, it has weakly higher profit and weakly higher welfare than $\menu$.

\xhdr{Strictness.}
Suppose that $\demdist(\quality)<\supdist(\quality)$ at some positive quality $\quality$.
The provider types $\costpar$ with $\demdist(\quality)<\costdis(\costpar)<\supdist(\quality)$ have positive probability and are positive, because $\costpar>\inversecostdis(\demdist(\quality))\geq\costL\geq0$.
Each of them selects a quality of at least $\quality$ under $\menu$ but, by \cref{eq:provider_threshold}, a lower quality under the replacement.
Because $\effort$ is strictly increasing, the expected production cost, and with it the expected provider compensation, is then strictly lower under the replacement, so both comparisons are strict.

\xhdr{Finite posted sets.}
Suppose that $\posted$ is finite.
Since $\qC$ takes values in $\posted$, the function $\demdist$ is constant on each interval between consecutive qualities of $\posted$, including its right endpoint, and vanishes above the largest quality of $\posted$.
Hence, for every level in $(0,1]$, the positive qualities at which $\demdist$ is at least this level form either the empty set or an interval whose right endpoint lies in $\posted$.
At every consumer type below $\valU$ and every provider type above $\costL$, the levels $1-\valdis(\val)$ and $\costdis(\costpar)$ are positive, so the choice maps constructed in the proof of \Cref{lem:implementation}, which take the supremum of such a set or zero if it is empty, select qualities in $\posted$.
Posting only the qualities in $\posted$, with the same price and wage at each of them, keeps these choices maximizing and changes choices only at the types $\valU$ and $\costL$.
The resulting menu posts exactly $\posted$ and has the same demand and supply distributions, profit, and welfare as the replacement.
\end{proof}
\subsection{Proof of \Cref{thm:commonconvexification}}\label{apx_commonconvexification}

\begin{lemma}\label{lemma:optmonotonicity}
For either scalar objective $\profitctrib(\cdot;\slope)$ or $\welfarectrib(\cdot;\slope)$, the largest maximizer over $\testmass\in[0,1]$ exists and is nonincreasing in the marginal technological cost $\slope>0$.
\end{lemma}

\begin{proof}
    Under bounded type supports, both scalar objectives are continuous on $[0,1]$, so their largest maximizers exist.

    Fix $0<\slope_1<\slope_2$, and let the trading masses $\testmass_1,\testmass_2$ be the largest maximizers of $\profitctrib(\cdot;\slope_1)$ and $\profitctrib(\cdot;\slope_2)$, respectively. By optimality,
    \begin{align*}
        \profitctrib(\testmass_1;\slope_1)\geq\profitctrib(\testmass_2;\slope_1), \quad
        \profitctrib(\testmass_2;\slope_2)\geq\profitctrib(\testmass_1;\slope_2).
    \end{align*}
    Adding these inequalities yields
    \begin{align*}
        (\slope_2-\slope_1)\cdot
        \bigl[\testmass_1\cdot\inversecostdis(\testmass_1)-\testmass_2\cdot\inversecostdis(\testmass_2)\bigr]\geq0.
    \end{align*}
    Since $\testmass\cdot\inversecostdis(\testmass)$ is strictly increasing, we obtain $\testmass_1\geq\testmass_2$.

    For welfare, the same argument applies with $\testmass\cdot\inversecostdis(\testmass)$ replaced by $\int_0^{\testmass}\inversecostdis(\massvar)\,\dd\massvar$, which is also strictly increasing. Hence the largest welfare maximizer is likewise nonincreasing in $\slope$.
\end{proof}

\begin{proof}[Proof of \Cref{thm:commonconvexification}]
    We first show that replacing $\effort$ with its greatest convex minorant $\convexeffort$ cannot decrease either objective for any fixed trading distribution.
    We then construct optimal trading distributions under $\convexeffort$ and show that they attain the same values under the original cost.

    \xhdr{Cost comparison.}
    The convexified cost $\convexeffort$ is continuous, strictly increasing, and absolutely continuous, and it agrees with the original cost at zero and at $\qualityU$.
    Let the cost gap be $\costgap(\quality)\deq\effort(\quality)-\convexeffort(\quality)$.
    By the properties of the greatest convex minorant, $\costgap$ is continuous and nonnegative, with $\costgap(0)=\costgap(\qualityU)=0$. Moreover, $\convexeffort$ is affine on every connected component of $\{\quality:\costgap(\quality)>0\}$, whose endpoints are contact qualities.

    Fix any trading distribution $\var$. Since $\var(\quality)$ is nonincreasing in $\quality$, and the function $x\inversecostdis(x)$ is nondecreasing in $x$, we have $B(\quality)=\var(\quality)\inversecostdis(\var(\quality))$.
    is nonincreasing in $\quality$. 
    By \cref{eq:profit-reduced}, since $B$ is nonincreasing and $\costgap(\quality)\geq0$ with $\costgap(0)=\costgap(\qualityU)=0$, integration by parts for monotone functions yields
    \begin{align*}
        \profit_{\convexeffort}(\var) - \profit_{\effort}(\var)
        =\int_0^{\qualityU}\costgap'(\quality)\cdot B(\quality)\,\dd \quality
        =-\int_0^{\qualityU}\costgap(\quality)\,\dd B(\quality)\geq0.
    \end{align*} 
    It follows that $\profit_{\effort}(\var)\leq\profit_{\convexeffort}(\var)$.
    For social welfare, $K(\quality)=\int_0^{\var(\quality)}\inversecostdis(\massvar)\,\dd\massvar$
    is also nonincreasing in $\quality$. 
    Applying the same argument gives $\welfare_{\effort}(\var)\leq\welfare_{\convexeffort}(\var)$.
    Besides, both inequalities become equalities whenever $\var$ is constant on every noncontact interval satisfying $\effort(\quality) > \convexeffort(\quality)$. 
    Indeed, in this case $B$ and $K$ are constant on these intervals, so their distributional derivatives are supported on the contact set, where $\costgap=0$.

    \xhdr{Optimal trading distributions.}
    Since $\convexeffort$ is convex and absolutely continuous, its derivative $\convexslope(\quality)$ exists at all but countably many qualities and is nondecreasing, and the two objectives under the convexified cost can be expressed as
    \begin{align*}
        \profit_{\convexeffort}(\var)
        &=\int_0^{\qualityU}\profitctrib(\var(\quality);\convexslope(\quality))\,\dd\quality, \quad
        \welfare_{\convexeffort}(\var)
        =\int_0^{\qualityU}\welfarectrib(\var(\quality);\convexslope(\quality))\,\dd\quality.
    \end{align*}
    For either objective, select the largest maximizer of its scalar contribution at marginal technological cost $\convexslope(\quality)$, at every quality $\quality$ where this derivative exists. By \Cref{lemma:optmonotonicity}, these maximizing masses are nonincreasing in $\slope$. Since $\convexslope$ is nondecreasing, selected masses are nonincreasing in quality on a set whose complement is countable. Taking their left-continuous versions gives trading distributions $\profitopt$ and $\welfareopt$, which agree with selected masses almost everywhere.

    These functions maximize respective scalar contributions almost everywhere. Thus, for any trading distribution $\var$, we have
    \begin{align*}
        \profit_{\convexeffort}(\var)
        &\leq\int_0^{\qualityU}\max_{x\in[0,1]}\profitctrib(x;\convexslope(\quality))\,\dd\quality
        =\profit_{\convexeffort}(\profitopt),\\
        \welfare_{\convexeffort}(\var)
        &\leq\int_0^{\qualityU}\max_{x\in[0,1]}\welfarectrib(x;\convexslope(\quality))\,\dd\quality
        =\welfare_{\convexeffort}(\welfareopt).
    \end{align*}
    Thus $\profitopt$ and $\welfareopt$ attain optimal values under the convexified cost.

    It remains to show that these optima are also attained under the original cost. On every noncontact interval, $\convexeffort$ is affine and hence $\convexslope(\quality)$ is constant. Because we consistently select the largest scalar maximizer, both $\profitopt$ and $\welfareopt$ are constant on each such interval. Their induced quality distributions therefore assign no mass to noncontact qualities.

    By the equality condition in the cost comparison,
    \begin{align*}
        \profit_{\effort}(\profitopt)=\profit_{\convexeffort}(\profitopt), \qquad
        \welfare_{\effort}(\welfareopt)=\welfare_{\convexeffort}(\welfareopt).
    \end{align*}
    Combining these equalities with the inequalities for a fixed trading distribution yields
    \begin{align*}
        \max_{\var}\profit_{\effort}(\var)
        =\max_{\var}\profit_{\convexeffort}(\var), \qquad
        \max_{\var}\welfare_{\effort}(\var)
        =\max_{\var}\welfare_{\convexeffort}(\var).
    \end{align*}
    Finally, \Cref{lem:implementation} implements both selected trading distributions with the bounded transfers in \cref{eq:price,eq:compensation} under the original technological cost.
    Since their induced quality distributions are supported on the contact qualities, the posted menus can be restricted to these qualities and the outside option without changing the induced quality distributions or objectives.

    Therefore, both optimal values are preserved and attained by menus supported on contact qualities.
\end{proof}

\subsection{Proof of \Cref{prop:finite_optimal_menu}}\label{apx:finite_menus}
\begin{proof}[Proof of \Cref{prop:finite_optimal_menu}]
For necessity, replace the transfers of a finite optimal menu by those in \cref{eq:price,eq:compensation} and delete unused offers.
Its original objective is no greater than its convexified objective, which is no greater than the integral of the scalar maxima.
\Cref{thm:commonconvexification} says the last value is the unrestricted original optimum.
Both inequalities must therefore be equalities.
For a finitely supported trading distribution with active qualities $\quality_1<\cdots<\quality_k$, trading masses $\testmass_1>\cdots>\testmass_k>0$, and $\testmass_{k+1}=0$, define $B(\testmass)\deq\testmass\cdot\inversecostdis(\testmass)$.
Integration by parts expresses the original-minus-minorant expenditure gap as
\begin{align*}
\sum_{i=1}^{k}
[\effort(\quality_i)-\convexeffort(\quality_i)]\cdot[B(\testmass_i)-B(\testmass_{i+1})].
\end{align*}
Each coefficient drop is strictly positive at an active quality, by strict increase of $\testmass\cdot\inversecostdis(\testmass)$.
Every summand is nonnegative, so equality forces every active quality to be a contact.
The pointwise profit gap is nonnegative and has integral zero; it therefore vanishes almost everywhere, which is exactly \cref{eq:finite_profit_criterion}.
For welfare, the coefficient is $\int_0^{\testmass}\inversecostdis(\massvar)\,\dd\massvar$, whose active drops are also strictly positive; the same two equality arguments apply.

For sufficiency, a finitely supported trading distribution satisfying the stated scalar condition attains the convexified integral maximum.
Because its active qualities are contacts, the original and convexified objectives are equal.
Bounded support makes every transfer in \cref{eq:price,eq:compensation} finite, including at full mass.
\Cref{lem:implementation} constructs its implementation, and deleting unused offers leaves a finite optimal posted menu.
\end{proof}

\section{Omitted Proofs in Section \ref{sec:efficiency}.}
\label{apx:efficiency}
The proofs in this section and in \Cref{apx:limitedmenus} use the marginal surplus $\commonfunc(\testmass;\slope)\deq\inversevaldis(1-\testmass)-\slope\cdot\inversecostdis(\testmass)$, so that $\profitctrib(\testmass;\slope)=\testmass\cdot\commonfunc(\testmass;\slope)$ and $\welfarectrib(\testmass;\slope)=\int_0^{\testmass}\commonfunc(\massvar;\slope)\,\dd\massvar$.

\subsection[Weak budget balance]{Proof of \Cref{prop:weak_budget_balance}}
\label{apx:wbb}

\begin{proof}[Proof of \Cref{prop:weak_budget_balance}]
    We first show nonnegative scalar profit at a welfare-maximizing trading mass, and then integrate this inequality along the convexified technological cost.
    
    Consider any marginal technological cost $\slope>0$ and scalar trading mass $\testmass\in[0,1]$.
    Since $\inversevaldis(1-\testmass)$ is nonincreasing and $\inversecostdis(\testmass)$ is nondecreasing, the marginal surplus $\commonfunc(\testmass;\slope)$ is nonincreasing and thus the scalar welfare contribution $\welfarectrib(\testmass;\slope)$ is concave.
    Let the scalar mass $\massW$ denote the largest maximizers of $\welfarectrib(\testmass;\slope)$.
    If $0<\massW<1$, optimality gives zero marginal surplus, i.e., $\commonfunc(\massW;\slope)=0$.
    If  $\massW=1$, the left derivative condition at this upper boundary gives $\commonfunc(1;\slope)\geq0$.
    Finally, if $\massW=0$, its profit contribution is $\profitctrib(0;\slope)=0$.
    Thus in every case the scalar profit contribution satisfies $\profitctrib(\massW;\slope)\geq0$.

    By the construction in \Cref{thm:commonconvexification}, the selected welfare-maximizing trading distribution $\welfareopt$ is constant on every noncontact interval.
    The equality condition in its proof therefore gives the same profit under the original and convexified costs.
    Moreover, $\welfareopt(\quality)$ is a scalar welfare maximizer at marginal technological cost $\convexeffort'_+(\quality)$ for almost every quality $\quality$, so we have
    \begin{align*}
        \profit(\welfareopt)
        =\profit_{\convexeffort}(\welfareopt)
        =\int_0^{\qualityU}\profitctrib\bigl(\welfareopt(\qualityvar);\convexeffort'_+(\qualityvar)\bigr)\,\dd\qualityvar
        \geq0,
    \end{align*}
    where the inequality follows from the nonnegative scalar profit established above.
    By \Cref{lem:implementation}, the transfers in \cref{eq:price,eq:compensation} implement $\welfareopt$ under the original cost, yielding a welfare-optimal menu with nonnegative platform profit.
\end{proof}

\subsection[Regularity counterexample]{Proof of \Cref{prop:regular_insufficient}}
\label{apx:regular_insufficient}

\begin{proof}[Proof of \Cref{prop:regular_insufficient}]
    We verify regularity, determine the scalar optima, and then use the continuous-quality reduction to show that these optima are attained against all menus.
    The provider expenditure curve is $\testmass\cdot\inversecostdis(\testmass)=\testmass^2$ and the consumer revenue curve is
    \begin{align*}
        \testmass\cdot\inversevaldis(1-\testmass)
        =\begin{dcases}
            \frac{\testmass}{\parameter}-\testmass^2,
            &0\leq\testmass\leq\parameter,\\
            (1+\parameter)\cdot(1-\testmass),
            &\parameter<\testmass\leq1.
        \end{dcases}
    \end{align*}
    The first consumer piece is concave and the second is linear.
    At the joining mass the revenue curve is continuous and its slope decreases, since $1/\parameter-2\cdot\parameter>-(1+\parameter)$.
    The provider expenditure curve is convex. 
    Hence both distributions are regular.

    Since the marginal technological cost is $\slope=1$, the scalar profit contribution is
    \begin{align*}
        \profitctrib(\testmass;1)
        &=\begin{dcases}
            \frac{\testmass}{\parameter}-2\cdot\testmass^2,
            &0\leq\testmass\leq\parameter,\\
            (1+\parameter)\cdot(1-\testmass)-\testmass^2,
            &\parameter<\testmass\leq1.
        \end{dcases}
    \end{align*}
    Since $0<\parameter<1/2$, the first piece is strictly increasing and the second is strictly decreasing.
    Thus the scalar profit-maximizing trading mass is uniquely $\massP=\parameter$.

    The welfare contribution is
    \begin{align*}
        \welfarectrib(\testmass;1)
        &=\begin{dcases}
            \displaystyle\int_0^{\testmass}\left(\frac{1}{\parameter}-2\cdot\massvar\right)\dd\massvar,
            &0\leq\testmass\leq\parameter,\\
            \displaystyle1-\parameter^2+\int_{\parameter}^{\testmass}
            \left((1+\parameter)\cdot\frac{1-\massvar}{\massvar}-\massvar\right)\dd\massvar,
            &\parameter<\testmass\leq1.
        \end{dcases}
    \end{align*}
    For $\parameter < 1/2$, the first piece is increasing on $[0,\parameter]$.
    Therefore the scalar welfare-maximizing mass $\massW\geq\parameter$ with
    \begin{align*}
        (\massW)^2&=(1+\parameter)\cdot(1-\massW),\\
        \massW&=\frac{\sqrt{(1+\parameter)\cdot(5+\parameter)}-(1+\parameter)}{2}
        \longrightarrow\frac{\sqrt5-1}{2}
        \quad\text{as }\parameter\to0.
    \end{align*}

    Because the technological cost is linear, \Cref{thm:commonconvexification} selects constant trading masses $\profitopt(\quality)=\massP$ and $\welfareopt(\quality)=\massW$ at every positive quality $\quality\leq1$.
    Their attained welfare values satisfy
    \begin{align*}
        \welfare(\profitopt)&=1-\parameter^2,\\
        \welfare(\welfareopt)
        &=1-\parameter^2+(1+\parameter)\cdot\ln\frac{\massW}{\parameter}\\
        &\quad-(1+\parameter)\cdot(\massW-\parameter)
        -\frac{(\massW)^2-\parameter^2}{2}.
    \end{align*}
    Since the welfare-maximizing mass converges to a positive constant, the welfare ratio $\welfare(\profitopt)/\welfare(\welfareopt)$ tends to zero as $\parameter$ tends to zero.
\end{proof}

\subsection[Profit versus welfare]{Proof of \Cref{thm:profit_vs_welfare}}
\label{apx_profit_vs_welfare}

\begin{proof}[Proof of \Cref{thm:profit_vs_welfare}]
    By \Cref{thm:commonconvexification}, it suffices to compare the largest profit and welfare maximizers at a fixed positive marginal technological cost, and apply the comparison along the convexified cost.
    
    Consider any marginal technological cost $\slope>0$ and scalar trading mass $\testmass\in[0,1]$.
    Let the scalar masses $\massP$ and $\massW$ denote the largest maximizers of the scalar profit and welfare contributions $\profitctrib(\testmass;\slope)$ and $\welfarectrib(\testmass;\slope)$, respectively.
    If the profit-maximizing mass is $\massP=0$, we have $\massP\leq\massW$.
    Otherwise, optimality implies $\profitctrib(\massP;\slope)\geq0$ and hence $\commonfunc(\massP;\slope)\geq0$.
    $\commonfunc(\testmass;\slope)$ is therefore nonnegative throughout the trading mass interval $[0,\massP]$ since it is nonincreasing.
    For any two trading masses $0\leq\masssmall\leq\masslarge\leq\massP$, this gives
    \begin{align*}
        \welfarectrib(\masslarge;\slope)-\welfarectrib(\masssmall;\slope)
        &\overset{(a)}{=}\int_{\masssmall}^{\masslarge}\commonfunc(\massvar;\slope)\dd\massvar
        \overset{(b)}{\geq}0,
    \end{align*}
    where equality~(a) holds by the definition of $\welfarectrib(\testmass;\slope)$ and 
    inequality~(b) holds by nonnegative marginal surplus on this interval.
    Thus $\welfarectrib(\testmass;\slope)$ is nondecreasing on $[0,\massP]$.
    The largest welfare maximizer $\massW$ therefore satisfies $\massP\leq\massW$.

    By the construction in \Cref{thm:commonconvexification}, $\profitopt(\qualityvar)$ and $\welfareopt(\qualityvar)$ are the largest scalar profit and welfare maximizers at marginal technological cost $\convexslope(\qualityvar)$ for almost every quality.
    Applying the above scalar comparison gives $\profitopt(\qualityvar)\leq\welfareopt(\qualityvar)$ for almost every $\qualityvar\in(0,\qualityU)$.
    Since both functions are left-continuous, approaching any positive quality from below through points where this inequality holds yields $\profitopt(\quality)\leq\welfareopt(\quality)$ for any $\quality\in(0,\qualityU]$.
\end{proof}

\subsection[Monotone hazard rate guarantee]{Proof of \Cref{thm:mhr_ratio}}
\label{apx:mhr_ratio}

\begin{example}\label{example:MHR_tight}
    Let $\qualityU=1$ and  $\effort(\quality)=\quality$ throughout the feasible interval $\quality\in[0,1]$.
    Thus the marginal technological cost is $\slope=1$.
    Fix a parameter $0<\parameter<1/2$.
    Define the consumer and provider distributions through quantile functions as $\inversecostdis(\testmass)=\parameter^2\cdot\testmass$ for $\testmass\in[0,1]$ and
    \begin{align*}
        \inversevaldis(1-\testmass)
        &\deq\begin{dcases}
            1+\frac{\parameter^2}{\ee}+\parameter\cdot(1-\ee\cdot\testmass),
            &0\leq\testmass\leq\frac{1}{\ee},\\
            \frac{\parameter^2}{\ee}-\ln\testmass,
            &\frac{1}{\ee}<\testmass\leq1.
        \end{dcases}
    \end{align*}
    The support of $\valdis$ is $[\parameter^2/\ee,1+\parameter+\parameter^2/\ee]$ and the support of $\costdis$ is $[0,\parameter^2]$.
    Both distributions are absolutely continuous and strictly increasing.
\end{example}

\begin{proof}[Proof of \Cref{thm:mhr_ratio}]
    We first prove the scalar ratio bound and integrate it along the convexified cost, and then show tightness by analyzing \Cref{example:MHR_tight}.

    \xhdr{Ratio bound.}
    Consider any positive marginal technological cost $\slope$ and scalar trading mass $\testmass\in[0,1]$.
    Define $\hazard(\testmass;\slope)\deq-\testmass\cdot\commonfunc'(\testmass;\slope)$.
    Under MHR conditions, derivatives below are understood almost everywhere:
    \begin{align*}
        -\testmass\cdot\frac{\dd}{\dd\testmass}\inversevaldis(1-\testmass)
        &=\frac{1}{\valhrate_{\valdis}(\inversevaldis(1-\testmass))},\\
        \testmass\cdot\frac{\dd}{\dd\testmass}\inversecostdis(\testmass)
        &=\frac{1}{\costhrate_{\costdis}(\inversecostdis(\testmass))}.
    \end{align*}
    Consequently, we have
    \begin{align*}
        \hazard(\testmass;\slope)
        =\frac{1}{\valhrate_{\valdis}(\inversevaldis(1-\testmass))}
        +\frac{\slope}{\costhrate_{\costdis}(\inversecostdis(\testmass))}.
    \end{align*}
    Since $\inversevaldis(1-\testmass)$ is nonincreasing and $\valhrate_{\valdis}$ is nondecreasing, the first term is nondecreasing in $\testmass$. 
    Similarly, since $\inversecostdis(\testmass)$ is nondecreasing and $\costhrate_{\costdis}$ is nonincreasing, the second term is also nondecreasing. 
    Hence, $\hazard(\testmass;\slope)$ is nondecreasing with $\hazard(\testmass+;\slope)$ denoting its right-hand limit.

    Let the scalar masses $\massP$ and $\massW$ denote the largest maximizers of the scalar profit and welfare contributions $\profitctrib(\testmass;\slope)$ and $\welfarectrib(\testmass;\slope)$, respectively.
    If $\welfarectrib(\massW;\slope)=0$, continuity and nonincrease of $\commonfunc(\testmass;\slope)$ imply that it is nonpositive everywhere.
    Thus $\massP=\massW=0$ and the desired bound is immediate.
    Otherwise $\commonfunc(\testmass;\slope) > 0$ at some $\testmass>0$, so $\profitctrib(\massP;\slope)>0$.
    Consequently $\massP>0$ and $\commonfunc(\massP;\slope)>0$.
    Recall that \Cref{thm:profit_vs_welfare} gives $\massP\leq\massW$.
    If $\massP=\massW$, the desired welfare bound is immediate.
    Consider the remaining case $0<\massP<\massW\leq1$, where $\massP$ is interior with
    \begin{align*}
        0&\overset{(a)}{\geq}\profitctrib'_{+}(\massP;\slope)
        \overset{(b)}{=}\commonfunc(\massP;\slope)-\hazard(\massP+;\slope),
    \end{align*}
    where inequality~(a) holds by profit optimality and equality~(b) differentiates the scalar profit contribution.
    For every $\testmass>\massP$, monotonicity of $\hazard(\testmass;\slope)$ gives $\hazard(\testmass;\slope) \geq \hazard(\massP+;\slope) \geq \commonfunc(\massP;\slope)$.
    Integrating the derivative then gives
    \begin{align*}
        \commonfunc(\testmass;\slope)
        =\commonfunc(\massP;\slope)
        -\int_{\massP}^{\testmass}\frac{\hazard(\massvar;\slope)}{\massvar}\dd\massvar
        \leq\commonfunc(\massP;\slope)\cdot
        \left(1-\ln\frac{\testmass}{\massP}\right).
    \end{align*}
    Since $\welfarectrib(\testmass;\slope)$ is maximized at $\massW$ satisfying $\commonfunc(\massW;\slope) \geq 0$ and $\commonfunc(\testmass;\slope)$ is nonincreasing, we can obtain $\commonfunc(\testmass;\slope) \geq 0$ for any $\testmass \leq \massW$.
    It follows that for any $\testmass \in [\massP, \massW]$,
    \begin{align*}
        0 \leq \commonfunc(\massW;\slope) \leq \commonfunc(\testmass;\slope) \leq \commonfunc(\massP;\slope)\cdot\left(1 - \ln \frac{\testmass}{\massP}\right).
    \end{align*}
    Since $\commonfunc(\massP;\slope) > 0$, we obtain $\massW/\massP\leq \ee$.
    Therefore, we have
    \begin{align*}
        \welfarectrib(\massW;\slope)-\welfarectrib(\massP;\slope)
        &=\int_{\massP}^{\massW}\commonfunc(\massvar;\slope)\dd\massvar\\
        &\leq\commonfunc(\massP;\slope)\cdot
        \int_{\massP}^{\ee\cdot\massP}\left(1-\ln\frac{\massvar}{\massP}\right)\dd\massvar\\
        &=(\ee-2)\cdot\massP\cdot\commonfunc(\massP;\slope).
    \end{align*}
    Monotonicity of $\commonfunc(\cdot;\slope)$ also gives
    \begin{align*}
        \welfarectrib(\massP;\slope)
        =\int_0^{\massP}\commonfunc(\massvar;\slope)\dd\massvar
        \geq\massP\cdot\commonfunc(\massP;\slope).
    \end{align*}
    Combining the last two bounds gives the scalar inequality 
    \begin{align*}
        \welfarectrib(\massP;\slope)\geq\welfarectrib(\massW;\slope)/(\ee-1)
    \end{align*}

    By \Cref{thm:commonconvexification}, the scalar inequality applies almost everywhere at marginal technological cost $\convexslope(\qualityvar)$, and the welfare values of $\profitopt$ and $\welfareopt$ are preserved under convexification.
    Hence, we can obtain that
    \begin{align*}
        \welfare(\profitopt)
        &{}=\int_0^{\qualityU}
        \welfarectrib\bigl(\profitopt(\qualityvar);\convexslope(\qualityvar)\bigr)\dd\qualityvar\\
        &{}\geq\frac{1}{\ee-1}\cdot\int_0^{\qualityU}
        \welfarectrib\bigl(\welfareopt(\qualityvar);\convexslope(\qualityvar)\bigr)\dd\qualityvar
        =\frac{1}{\ee-1}\cdot\welfare(\welfareopt).
    \end{align*}

    \xhdr{Tightness.}
    We now show the tightness by analyzing \Cref{example:MHR_tight}.
    The consumer distribution satisfies
    \begin{align*}
        -\testmass\cdot\frac{\dd}{\dd\testmass}\inversevaldis(1-\testmass)
        =\frac{1}{\valhrate_{\valdis}(\inversevaldis(1-\testmass))}
        &=\begin{dcases}
            \parameter\cdot\ee\cdot\testmass,&0<\testmass<\frac1\ee,\\
            1,&\frac1\ee<\testmass<1.
        \end{dcases}
    \end{align*}
    One can verify that this quantity is nondecreasing in $\testmass$, implying $\valhrate_{\valdis}$ is nondecreasing.
    The provider distribution $\costdis$ has reverse hazard $\costhrate_{\costdis}(\costpar)=1/\costpar$, which is nonincreasing.
    Thus $\valdis$ and $\costdis$ satisfy MHR.

    Since the marginal technological cost is $\slope=1$, the marginal surplus is
    \begin{align*}
        \commonfunc(\testmass;1)
        &=\begin{dcases}
            1+\frac{\parameter^2}{\ee}+\parameter\cdot(1-\ee\cdot\testmass)-\parameter^2\cdot\testmass,
            &0\leq\testmass\leq\frac1\ee,\\
            -\ln\testmass+\parameter^2\cdot\left(\frac1\ee-\testmass\right),
            &\frac1\ee<\testmass\leq1.
        \end{dcases}
    \end{align*}
    A direct differentiation shows that the scalar profit contribution $\profitctrib(\testmass;\slope)$ is increasing on $(0,1/\ee)$ and decreasing on $(1/\ee,1)$.
    Thus the unique scalar profit-maximizing mass is $\massP=1/\ee$.

    Note that 
    \begin{align*}
        \commonfunc\left(\frac1\ee;1\right) = 1, 
        \qquad \commonfunc(1;1) =\parameter^2\cdot\left(\frac1\ee-1\right) < 0.
    \end{align*}
    The unique scalar welfare-maximizing mass $\massW\in(1/\ee,1)$ satisfies
    \begin{align*}
        0<
        -\ln\massW 
        = \parameter^2\cdot\left(\massW-\frac1\ee\right)
        <\parameter^2\cdot\left(1-\frac1\ee\right).
    \end{align*}
    Thus $\massW$ tends to $1$ as $\parameter$ tends to $0$.

    Since $\effort(\quality) = \quality$, \Cref{thm:commonconvexification} implies that both optima are attained by menus offering only quality $1$ and the outside option. 
    Their welfare values satisfy
    \begin{align*}
        \welfare(\profitopt)&=\frac{1+\parameter/2}{\ee}+\frac{\parameter^2}{2\cdot\ee^2},\\
        \welfare(\welfareopt)&=\welfare(\profitopt)+\int_{1/\ee}^{\massW}
        \left(-\ln\massvar+\parameter^2\cdot\left(\frac1\ee-\massvar\right)\right)\dd\massvar.
    \end{align*}
    As $\parameter$ tends to $0$, $\welfare(\profitopt)$ tends to $1/\ee$ and thus $\welfare(\welfareopt)$ tends to $(\ee-1)/\ee$.
    Hence the welfare ratio tends to $1/(\ee-1)$ as desired.

\end{proof}

\section{Omitted Proofs in Section \ref{sec:limitedmenus}.}
\label{apx:limitedmenus}
\subsection[Tight finite-menu guarantee]{Proof of \Cref{thm:restrictmenu_profitloss}}
\label{app:limitedmenus_continuous}

\begin{example}\label{example:limited_menu_tight}
    Fix an integer $\tiers\geq1$ and a parameter $\rho>1$.
    Both type distributions are uniform on $[0,1]$.
    Let $\qualityU=1$ and define the normalization constant $\normconst\deq\sum_{\tierindex=1}^{\tiers}\rho^{\tierindex}$.
    Define the quality nodes by
    \begin{align*}
        \qTier{0}\deq0, \quad
        \qTier{\tierindex}\deq\frac{1}{\normconst}\cdot\sum_{\sumindex=1}^{\tierindex}\rho^{\sumindex},
        \qquad 1\leq\tierindex\leq\tiers.
    \end{align*}
    Thus $0=\qTier{0}<\qTier{1}<\cdots<\qTier{\tiers}=1$.
    Define 
    \begin{align*}
        \effort(\qTier{\tierindex})
        \deq\frac{1}{\normconst}\cdot\sum_{\sumindex=1}^{\tierindex}
            \left(\rho^{2\cdot\sumindex}-\rho^{\sumindex}\right),
        \qquad 1\leq\tierindex\leq\tiers,
    \end{align*}
    and let $\effort$ be linear on each interval $[\qTier{\tierindex-1},\qTier{\tierindex}]$ with slope $\rho^{\tierindex}-1$.
    These slopes are positive and strictly increasing, so $\effort$ is absolutely continuous, strictly increasing, and convex.
\end{example}

\begin{proof}[Proof of \Cref{thm:restrictmenu_profitloss}]
    We first prove the profit guarantee by deleting qualities, and then establish tightness using \Cref{example:limited_menu_tight}.

    \xhdr{The guarantee.}
    Choose an optimal menu whose transfers are given by \cref{eq:price,eq:compensation}, with $\tiers$ active positive qualities $0=\qTier{0}<\qTier{1}<\cdots<\qTier{\tiers}$ and trading masses $\testmass_1>\cdots>\testmass_{\tiers}>0$.
    Define its quality increments, cost increments, and slopes by
    \begin{align*}
        \qualitygap_{\tierindex} \deq\qTier{\tierindex}-\qTier{\tierindex-1}, \quad
        \effortgap_{\tierindex}
        \deq\effort(\qTier{\tierindex}) -\effort(\qTier{\tierindex-1}), \quad
        \slope_{\tierindex}
        \deq\frac{\effortgap_{\tierindex}}{\qualitygap_{\tierindex}}.
    \end{align*}
    By the necessity argument in \Cref{prop:finite_optimal_menu}, its active qualities are contacts and its mass maximizes scalar profit almost everywhere under the convexified cost.
    Consequently, each incremental profit satisfies $\profit_{\tierindex}    \deq\qualitygap_{\tierindex}\cdot\profitctrib(\testmass_{\tierindex};\slope_{\tierindex}) \geq 0$.

    Now consider a menu whose transfers are given by \cref{eq:price,eq:compensation}, with $k$ active qualities, nonnegative incremental profits, and total profit $V_k$.
    Choose an increment $j$ with $\profit_j\leq V_k/k$.
    If $j=k$, delete the highest quality.
    Otherwise, lower $\testmass_j$ to $\testmass_{j+1}$.
    Since $\commonfunc(\cdot;\slope_j)$ is decreasing, we can obtain that
    \begin{align*}
        \commonfunc(\testmass_{j+1};\slope_j)
        \geq\commonfunc(\testmass_j;\slope_j)
        =\frac{\profit_j}{\qualitygap_j\cdot\testmass_j}
        \geq0.
    \end{align*}
    Hence the modified increment still contributes nonnegative profit, and the loss is at most $\profit_j$.
    This operation deletes quality $\qTier{j}$ from the menu.
    Thus the resulting menu has $k-1$ active qualities and
    \begin{align*}
        V_{k-1}\geq V_k-\profit_j
        \geq\frac{k-1}{k}V_k.
    \end{align*}
    Starting from $V_{\tiers}=\oprofit$ and repeating this deletion gives
    \begin{align*}
        \oprofit_{\menusize}
        \geq V_{\menusize}
        \geq\left(\prod_{k=\menusize+1}^{\tiers} \frac{k-1}{k}\right)\oprofit
        =\frac{\menusize}{\tiers}\oprofit.
    \end{align*}

    \xhdr{Tightness.}
    Consider \Cref{example:limited_menu_tight}.
    Uniform types give $\profitctrib(\testmass;\slope)=\testmass-(1+\slope)\cdot\testmass^2$.
    On increment $\tierindex$, we have $\profit_{\tierindex} = (\rho^{\tierindex}\cdot\testmass-\rho^{2\cdot\tierindex}\cdot\testmass^2)/\normconst$.
    Its unique maximizer is $\testmass_{\tierindex}=1/(2\cdot\rho^{\tierindex})$,
    with value $1/(4\cdot\normconst)$.
    These masses strictly decrease, so
    \Cref{thm:commonconvexification} gives $\oprofit=\tiers/(4\cdot\normconst)$.
    
    Uniqueness forces every optimal trading distribution to attain these values almost everywhere on the corresponding increments.
    By left continuity, all $\tiers$ increment endpoints are active, so every optimal menu contains exactly $\tiers$ positive qualities.
    
    For restricted menus, it suffices to consider increment endpoints. 
    Indeed, fixing the trading masses, profit is affine in each quality location within a linear cost increment. 
    Thus any interior quality can be moved to an increment endpoint or merged with an adjacent quality without decreasing profit or increasing menu size. 
    Repeating this process proves the claim.

    An endpoint menu with at most $\menusize$ active qualities partitions a prefix of the increments into at most $\menusize$ consecutive groups, followed by a zero-mass suffix.
    For a group of increments $a,\ldots,b$, its profit is at most
    \begin{align*}
        \max_{\testmass\in[0,1]}
            \frac{1}{\normconst}\cdot\sum_{\tierindex=a}^{b}
            \bigl(\rho^{\tierindex}\cdot\testmass
                -\rho^{2\cdot\tierindex}\cdot\testmass^2\bigr)
        &\leq
            \frac{\left(\sum_{\tierindex=a}^{b}
                \rho^{\tierindex}\right)^2}
            {4\cdot\normconst\cdot\sum_{\tierindex=a}^{b}\rho^{2\cdot\tierindex}}\\
        &=\frac{1}{4\cdot\normconst}\cdot\frac{\rho+1}{\rho-1}
            \cdot\frac{\rho^{b-a+1}-1}{\rho^{b-a+1}+1}
            \leq\frac{1}{4\cdot\normconst}\cdot\frac{\rho+1}{\rho-1}.
    \end{align*}
    Summing over at most $\menusize$ groups and applying the guarantee already proved yields $\menusize/\tiers \leq \oprofit_{\menusize}/\oprofit \leq \menusize/\tiers \cdot (\rho+1)/(\rho-1)$.
    Letting $\rho\to\infty$ proves tightness.
\end{proof}

\subsection[Profit and welfare under MHR]{Proof of \Cref{lem:mhr_profit_welfare}}
\label{app:mhr_profit_welfare}

\begin{proof}
    We first show that reducing a scalar trading mass by a factor of $\ee$ yields profit at least $1/\ee$ times its original welfare contribution.
    We then apply this comparison at every quality threshold and implement the reduced trading distribution.

    Fix a nonnegative marginal technological cost $\slope$.
    Consumer MHR makes the map $\val\mapsto-\log(1-\valdis(\val))$ increasing and convex, since its derivative is the nondecreasing consumer hazard rate.
    With $\logmass\leq0$ denoting logarithmic trading mass, the map $\logmass\mapsto\inversevaldis(1-\ee^{\logmass})$ is therefore concave.
    Similarly, the map $\logmass\mapsto\inversecostdis(\ee^{\logmass})$ is convex.
    It follows that $\logmass\mapsto\commonfunc(\ee^{\logmass};\slope)$ is concave for $\logmass\leq0$.

    For a positive trading mass $\testmass\in(0,1]$, we obtain
    \begin{align*}
        \frac{\welfarectrib(\testmass;\slope)}{\testmass}
        &=\frac{1}{\testmass}\cdot\int_0^{\testmass}\commonfunc(\massvar;\slope)\,\dd\massvar\\
        &\overset{(a)}{\leq}\commonfunc(\cdot;\slope)\!\left(
            \exp\!\left[\frac{1}{\testmass}\cdot\int_0^{\testmass}\log\massvar\,\dd\massvar\right]\right)
        =\commonfunc(\frac{\testmass}{\ee};\slope),
    \end{align*}
    where inequality~(a) follows from Jensen's inequality.
    Multiplying by $\testmass/\ee$ gives $\profitctrib(\testmass/\ee;\slope)
        \geq \welfarectrib(\testmass;\slope)/\ee$.
    The same inequality holds at $\testmass=0$, since both sides vanish.

    Since $\welfareopt/\ee$ is nonincreasing, left-continuous, and takes values in $[0,1]$, \Cref{lem:implementation} implements it with the transfers in \cref{eq:price,eq:compensation}.
    Applying the above inequality at the nonnegative marginal technological cost $\effort'(\quality)$ for almost every $\quality$ and integrating yields
    \begin{align*}
        \oprofit
        &\geq\profit(\welfareopt/\ee)
        =\int_0^{\qualityU}
            \profitctrib(\welfareopt(\quality)/\ee;\effort'(\quality))\,\dd\quality\\
        &\geq\frac{1}{\ee}\cdot\int_0^{\qualityU}
            \welfarectrib(\welfareopt(\quality);\effort'(\quality))\,\dd\quality
        =\frac{\welfare(\welfareopt)}{\ee}
    \end{align*}
\end{proof}

\subsection[Additive approximation and dynamic programming]{Proof of \Cref{thm:compute_restricted_menu}}
\label{app:restricted_dp_continuous}

\begin{proof}
    In this proof, we show the error, correctness, and complexity of our algorithm, respectively.
    
    \xhdr{Error analysis.}
    We bound two sources of error by successive downward rounding: menu-size restriction and computational grid. 
        Consider a strictly matched menu whose transfers are given by \cref{eq:price,eq:compensation}. By definition,
    $\Ucons(\val)=\val\qC(\val)-\price(\qC(\val))$ and
    $\Uprov(\costpar)=\wage(\qP(\costpar))-\costpar\effort(\qP(\costpar))$.
    Each indirect utility is the maximum of affine functions of its type.
    Since qualities and technological costs are bounded, the utilities
    are Lipschitz continuous. At almost every type, their derivatives
    equal the type coefficients of chosen offers: $\Ucons'(\val)=\qC(\val)$ and $\Uprov'(\costpar)=-\effort(\qP(\costpar))$.
    Integrating these derivatives and using endpoint
    utilities $\Ucons(\valL)=\Uprov(\costU)=0$ gives
    \begin{align*}
        \Ucons(\val)=\int_{\valL}^{\val}\qC(\typevar)\,\dd\typevar, \quad
        \Uprov(\costpar)
        =\int_{\costpar}^{\costU}
            \effort(\qP(\typevar))\,\dd\typevar.
    \end{align*}
    Taking expectations and exchanging the order of integration yields
    \begin{align*}
        \expect[\val]{\Ucons(\val)}
        &=\int_{\valL}^{\valU}\qC(\typevar)
            \left(\int_{\typevar}^{\valU}
                \valdens(\val)\,\dd\val\right)\dd\typevar\\
        &=\int_{\valL}^{\valU}
            (1-\valdis(\val))\qC(\val)\,\dd\val,\\
        \expect[\costpar]{\Uprov(\costpar)}
        &=\int_{\costL}^{\costU}\effort(\qP(\typevar))
            \left(\int_{\costL}^{\typevar}
                \costdens(\costpar)\,\dd\costpar\right)\dd\typevar\\
        &=\int_{\costL}^{\costU}
            \costdis(\costpar)\effort(\qP(\costpar))\,\dd\costpar.
    \end{align*}
    These exchanges are valid because the integrands are nonnegative
    and the integrals are finite under bounded qualities and costs.

    Recall that
    \begin{align*}
        \price(\qC(\val))
        &=\val\qC(\val)-\Ucons(\val),\\
        \wage(\qP(\costpar))
        &=\costpar\effort(\qP(\costpar))+\Uprov(\costpar).
    \end{align*}
    Taking expectations and substituting preceding expressions gives
    \begin{align*}
        \expect[\val]{\price(\qC(\val))}
        &=\int_{\valL}^{\valU}
            [\val\valdens(\val)-(1-\valdis(\val))]
            \qC(\val)\,\dd\val,\\
        \expect[\costpar]{\wage(\qP(\costpar))}
        &=\int_{\costL}^{\costU}
            [\costpar\costdens(\costpar)+\costdis(\costpar)]
            \effort(\qP(\costpar))\,\dd\costpar.
    \end{align*}

    For an integer $\truncation\geq1$, round qualities by $r_{\truncation}(\quality)=(\qualityU/\truncation)\lfloor\truncation\quality/\qualityU\rfloor$. 
    Applying this map on both sides preserves equality of their quality distributions and monotonicity in type, so \Cref{lem:implementation} implements the rounded choices with the transfers in \cref{eq:price,eq:compensation}.
    
    On provider side, since $\costpar\costdens(\costpar)+\costdis(\costpar) \geq 0$ and $\effort$ is increasing, we have $\expect[\costpar]{\wage(\qP(\costpar))}$ weakly decreases. 
    
    On consumer side, writing $e(\val)=\qC(\val)-r_{\truncation}(\qC(\val))\in[0,\qualityU/\truncation]$, the expected consumer payment loss is at most
    \begin{align*}
        \int_{\valL}^{\valU}
            [\val\valdens(\val)-(1-\valdis(\val))]e(\val)\,\dd\val\leq\int_{\valL}^{\valU}\val\valdens(\val)e(\val)\,\dd\val
        \leq\frac{\qualityU\consumermean}{\truncation}.
    \end{align*}
    
    Thus profit decreases by at most $\qualityU \cdot \consumermean/\truncation$.

    Start from an unrestricted optimum $\oprofit$, which exists by \Cref{thm:commonconvexification,thm:characterization}. 
    Rounding with $\truncation=\menusize$ gives at most $\menusize$ positive qualities and loses at most $\qualityU \cdot \consumermean/\menusize$. 
    Rounding this menu again with $\truncation=\approxstage$ places it on $\qualities_{\approxstage}$ without increasing its size, and loses at most $\qualityU \cdot \consumermean/\approxstage\leq\qualityU \cdot \valU/\approxstage\leq\approxrate$. 
    The transfers in \cref{eq:price,eq:compensation} are recomputed after each rounding. 
    Hence a feasible restricted grid menu has profit at least $\oprofit-(\qualityU \cdot \consumermean)/\menusize-\approxrate$.

    \xhdr{Algorithm correctness.}
    We first show that convexification preserves the optimal
    $\menusize$-restricted grid profit, and then verify the
    dynamic program.
    
    For convexification, let $\widehat{\effort}$ be the piecewise-linear curve in Step~1 and let $\underline{\widehat{\effort}}$ be its greatest convex minorant.
    By the proof of \Cref{thm:commonconvexification}, replacing $\widehat{\effort}$ by $\underline{\widehat{\effort}}$ weakly increases profit for any fixed trading distribution.

    Consider any feasible $\menusize$-restricted grid menu.
    Under $\underline{\widehat{\effort}}$, fix its trading masses and consider a noncontact quality $\quality$.
    Let $(a,b)$ be the maximal interval containing $\quality$ on which $\widehat{\effort}> \underline{\widehat{\effort}}$.
    Its endpoints are contact grid qualities, and
    $\underline{\widehat{\effort}}$ is affine on $[a,b]$.

    Let $\masslarge$ and $\masssmall$ be the trading masses immediately below and above $\quality$.
    With other quality locations fixed, the total platform profit under the convexified cost is
    \begin{align*}
        \profit_{\underline{\widehat{\effort}}}
        &=C+\quality\bigl[
            \masslarge\inversevaldis(1-\masslarge)
            -\masssmall\inversevaldis(1-\masssmall)\bigr]\\
        &\quad-\underline{\widehat{\effort}}(\quality)\bigl[
            \masslarge\inversecostdis(\masslarge)
            -\masssmall\inversecostdis(\masssmall)\bigr],
    \end{align*}
    where $C$ is independent of $\quality$.
    Hence $\profit_{\underline{\widehat{\effort}}}$ is also affine on $[a,b]$ implying that $\quality$ can move toward $a$ or $b$ without decreasing profit, until it reaches that endpoint or another offered quality. 
    In the latter case, merge the coincident offers and repeat the argument.
    Each move either reaches a contact or reduces the number of distinct interior qualities, so this process terminates without increasing menu size.

    Repeating for all noncontact qualities gives a contact menu with weakly higher convexified profit and at most $\menusize$ active positive qualities.
    Since $\widehat{\effort}=\underline{\widehat{\effort}}$ at every resulting quality, its profit under the original sampled costs is the same.
    Thus restricting to the retained contact qualities preserves the optimal $\menusize$-restricted grid profit.

    On the retained qualities, any nonzero trading distribution with at most $\menusize$ active qualities partitions a prefix into at most $\menusize$ constant-mass groups, followed by a zero suffix.
    For a group $(\qTier{\leftindex},\qTier{\rightindex}]$ with mass $\testmass$, its profit contribution satisfies $(\qTier{\rightindex}-\qTier{\leftindex})\cdot
            \profitctrib(\testmass;\slope_{\leftindex,\rightindex})
        \leq\profit(\leftindex,\rightindex)$,
    where $\slope_{\leftindex,\rightindex}$ and $\profit(\leftindex,\rightindex)$ are the group slope and optimal group profit defined in Step~3.
    Conversely, successive groups have nondecreasing slopes, so their largest maximizing masses $\oracle(\slope_{\leftindex,\rightindex})$ are nonincreasing by \Cref{lemma:optmonotonicity}.
    They therefore realize the sum of group values with at most one active quality per group.

    The problem is consequently equivalent to the stated consecutive-group partition problem.
    Conditioning on the last group's left endpoint gives the recurrence for the partition value $\dpfunc[\rightindex,\groupcount]$; induction on $\groupcount$ proves its optimality.
    Maximizing over all permitted prefixes and group counts, together with the zero-trade solution, yields the optimal restricted grid profit.
    Backtracking and \Cref{lem:implementation} implement this value with at most $\menusize$ active positive qualities.

    \xhdr{Welfare under MHR.}
    Suppose both type distributions satisfy MHR.
    The social welfare of the output menu $\menu_{\menusize,\approxrate}$ satisfies
    \begin{align*}
        \welfare(\menu_{\menusize,\approxrate})
        &\geq\profit(\menu_{\menusize,\approxrate})
        \overset{(a)}{\geq}\oprofit-\frac{\qualityU\cdot\consumermean}{\menusize}-\approxrate\\
        &\overset{(b)}{\geq}\frac{\welfare(\welfareopt)}{\ee}
            -\frac{\qualityU\cdot\consumermean}{\menusize}-\approxrate.
    \end{align*}
    Inequality~(a) is the profit guarantee established above.
    Inequality~(b) follows from \Cref{lem:mhr_profit_welfare}.

    \xhdr{Complexity.}
    The algorithm uses $\approxstage+1$ cost evaluations.
    Convexification takes $\bO{\approxstage}$ arithmetic operations.
    Since $\convextiers\leq\approxstage$, group values and maximizing masses require $\bO{\approxstage^2}$ oracle calls. 
    Endpoint differences give each group's increments in constant time. 
    There are $\bO{\min\{\menusize,\convextiers\}\convextiers}$ DP states and at most $\convextiers$ transitions per state, giving $\bO{\min\{\menusize,\approxstage\}\approxstage^2}$
    arithmetic operations in total. Recovering transfers uses $\bO{\approxstage}$ quantile evaluations and arithmetic operations.
\end{proof}

\end{document}